\documentclass[nonacm]{acmart}
\usepackage{booktabs}
\usepackage{graphicx}

\usepackage{subfig}
\usepackage{tikz}
\usetikzlibrary{shapes}
\usepackage{pgf}
\usepackage{amsfonts}
\usepackage{amsmath}
\usepackage{algorithm}
\usepackage{algorithmic}
\usepackage{xspace}

\newtheorem{exmp}{Example}[section]

\usepackage{multirow}
\usepackage{threeparttable}

\newcommand{\btitle}[1]{\vspace{.5ex}\noindent \textbf{#1}}
\newcommand{\eat}[1]{}
\newcommand{\tabincell}[2]{\begin{tabular}{@{}#1@{}}#2\end{tabular}}
\newcommand{\ie}{\emph{i.e.,}\xspace}
\newcommand{\eg}{\emph{e.g.,}\xspace}
\newcommand{\etal}{\emph{et al.}\xspace}

\AtBeginDocument{%
  \providecommand\BibTeX{{%
    \normalfont B\kern-0.5em{\scshape i\kern-0.25em b}\kern-0.8em\TeX}}}

\acmJournal{CSUR}

\begin{document}

\title{Influence Maximization in Social Networks: A Survey}





\author{Hui Li}
\author{Susu Yang}
\author{Mengting Xu}
\affiliation{%
  \institution{Xidian University}
  \streetaddress{Xifeng Road 266}
  \city{Xi'an}
  \country{China}
  \postcode{710126}}
\email{hli@xidian.edu.cn}

\author{Sourav S Bhowmick}
\affiliation{%
	\institution{Nanyang Technological University}
	\country{Singapore}}
\email{assourav@ntu.edu.sg}

\author{Jiangtao Cui}
\affiliation{%
	\institution{Xidian University}
	\country{China}}
%

\renewcommand{\shortauthors}{Li et al.}

\begin{abstract}
  Online social networks have become an important platform for people to communicate, share knowledge and disseminate information. Given the widespread usage of social media, individuals' ideas, preferences and behavior are often influenced by their peers or friends in the social networks that they participate in. Since the last decade, \textit{influence maximization} (\textsc{im}) problem has been extensively adopted to model the diffusion of innovations and ideas. The purpose of \textsc{im} is to select a set of $k$ seed nodes who can influence the most individuals in the network.
  In this survey, we present a systematical study over the researches and future directions with respect to \textsc{im} problem. We review the information diffusion models and analyze a variety of algorithms for the classic IM algorithms. We propose a taxonomy for potential readers to understand the key techniques and challenges. We also organize the milestone works in time order such that the readers of this survey can experience the research roadmap in this field. Moreover, we also categorize other application-oriented \textsc{im} studies and correspondingly study each of them. What's more, we list a series of open questions as the future directions for \textsc{im}-related researches, where a potential reader of this survey can easily observe what should be done next in this field.
\end{abstract}



\maketitle

\vspace{-1ex}\section{Introduction}
Platforms of online social networks (\eg \textit{Twitter}, \textit{Facebook}) have become increasingly popular since the last decade. They have been the source of massive volumes of social data and have given rise to several interesting real-world applications such as viral marketing, fake news detection, and information cascade prediction. Specifically, in viral marketing, a marketer may provide some individuals in a social network with free products in exchange for them to spread the good word about it. If this endeavor is successful, it may create a large cascade of influence on social network users via ``word-of-mouth'' recommendations. In this context, Kempe \etal ~\cite{DBLP:conf/kdd/KempeKT03} introduced the \textit{influence maximization} (IM) problem to investigate how these initial individuals (\ie \textit{seed} nodes) can be selected to maximize the final influence.

Formally, given a social network $G$ and a budget parameter $k$, the IM problem aims to find $k$ users to maximize the number of influenced users at the end of information diffusion. Kempe \etal \cite{DBLP:conf/kdd/KempeKT03} first formalized this problem and proved that finding the optimal solution for the problem is NP-hard. They presented an approximation algorithm based on hill climbing strategy that iteratively selects a node with maximum \textit{marginal influence}. This strategy is proven to guarantee a $(1-1/e)$ approximation ratio with respect to the optimal solution. However, as Monte-Carlo simulation is employed to estimate the marginal influence of nodes during each iteration, the time cost is prohibitive especially for large networks. Consequently, Kempe \etal \cite{DBLP:conf/kdd/KempeKT03} introduced a pair of heuristic solutions driven by node degree and PageRank, respectively, in order to strike a balance between result quality and efficiency. Since then, the research community has developed numerous approximate and heuristic approaches to improve the seminal solutions proposed by Kempe \etal Specifically, the approximate solutions aim to reduce the cost while guaranteeing the approximation ratio whereas the heuristic ones attempt to find a superior balance between efficiency and result quality in practice. Although the latter ones may often exhibit lower cost than the former ones, they fail to provide any theoretical guarantee on the result quality.

\vspace{-2ex}\subsection{Our Contributions}
In this survey, we present a comprehensive analysis of the IM problem; from definitions of related concepts, IM techniques, to various realistic extensions to the classical IM problem. Through the survey, readers can quickly understand and get into the area of influence maximization. Specifically, our contributions include the following. First, we introduce the IM problem formally along with the information diffusion models that underpin it. We present a taxonomy to organize existing solutions and illustrate the development pathway along the key solution categories. Second, we conduct an exhaustive study over key existing techniques for the classical IM problem, where we not only reveal the development trends of existing work but also highlight the strengths and weaknesses of each effort. Third, we discuss how the classical IM problem is extended to incorporate real world requirements. We also provide a systematic insights over \textit{adaptive} IM, a recent derivative of the classical IM problem.
Lastly, we articulate the challenges and open issues, and identify new trends and future directions in this arena.

There exist some experimental studies~\cite{DBLP:conf/sigmod/AroraGR17,DBLP:conf/sigmod/Ohsaka20} and surveys~\cite{DBLP:journals/fgcs/NBB18,DBLP:journals/tkde/LiFWT18,DBLP:journals/kais/BanerjeeJP20} on the IM problem. In particular, \cite{DBLP:conf/sigmod/AroraGR17,DBLP:conf/sigmod/Ohsaka20} present an experimental analysis over a number of representative IM efforts, focusing on experimental comparison between different solutions w.r.t seeds quality, time efficiency and memory consumption. It is, however, not a survey and hence does not provide any in-depth theoretical insight of the IM problem or techniques. On the other hand, some shallow surveys like \cite{DBLP:journals/fgcs/NBB18} and \cite{DBLP:journals/kais/BanerjeeJP20} categorize works only at the highest level, namely approximation and heuristic but no other detailed study inside the sub-levels within both categories. Besides, they only take into account the progressive diffusion models (IC, LT, and CT) but not non-progressive models. In addition, neither or them has provided any discussion on extended IM issues.

More germane to our work is the recent survey in~\cite{DBLP:journals/tkde/LiFWT18}. Our work differs in the following key ways. Firstly, we classify existing IM efforts into \textit{approximate} and \textit{heuristic} ones. In contrast, the criteria used for classification in~\cite{DBLP:journals/tkde/LiFWT18} is based on the assumption that IM techniques have good theoretical result guarantee (\ie approximation ratio). Consequently, it classifies the IM techniques into \textit{simulation-based} and \textit{sampling-based} approaches. However, both these approaches share the same building block, hill-climbing strategy, and only differentiates themselves during the estimation of influence for the candidate seeds. Furthermore, they are in fact two sides of the same coin. That is, both are under the category of approximation algorithms. Consequently, they ignore solutions that have no theoretical guarantee but good practical solutions. As more new heuristic IM techniques emerge, they cannot be accurately classified into any of the categories in~\cite{DBLP:journals/tkde/LiFWT18}. In addition, the claim that simulation-based approaches generally support many information diffusion models compared to sampling-based ones does not hold. Examples that contradict this claim, in fact, can be found in Table 1 of~\cite{DBLP:journals/tkde/LiFWT18} (3 out of the 6 simulation-based approaches contradict this claim \eg \textsf{UBLF}~\cite{DBLP:journals/tkde/ZhouZZG15}, \textsf{CGA}~\cite{DBLP:conf/kdd/WangCSX10}, \textsf{SA}~\cite{DBLP:conf/aaai/JiangSCWSX11}). Secondly, we present a different categorization for the information diffusion models by classifying them into \textit{progressive} and \textit{non-progressive} ones. This is consistent with a recent study on information diffusion models~\cite{DBLP:journals/sigmod/GuilleHFZ13}. However, such clear categorization is missing in~\cite{DBLP:journals/tkde/LiFWT18}. Thirdly, we survey adaptive \textsc{im} techniques that have received increasing attention recently. Discussions on this important variant of IM problem are missing in~\cite{DBLP:journals/tkde/LiFWT18}.

\vspace{-2ex}\subsection{Survey Organization}
The rest of this paper is organized as follows. We formally define the IM problem and introduce related concepts associated with it in Section~\ref{sec3}.  In Section~\ref{sec2}, we introduce the framework of our survey. Specifically, our framework is organized into the following three parts: (a) information diffusion models, (b) algorithms for classical IM problem, and (c) techniques for extended IM problem. We elaborate on each one of them in Sections~\ref{sec4}, \ref{sec5}, and \ref{sec6}, respectively. Section~\ref{sec7} discusses future directions of IM research. Finally, we conclude the survey in Section~\ref{sec8}.

\vspace{-1ex}\section{Preliminaries}\label{sec3}
In this section, we present the formal definition of the \textit{influence maximization} (IM) problem and related concepts to facilitate exposition of IM research in the literature. Key notations that are used throughout this survey are given in Table~\ref{tab:variables}.

\begin{table}[t]
	\caption{Key notations used in the paper.}\vspace{-2ex}
	\scriptsize
	\label{tab:variables}
	\begin{tabular}{c|l}
		\toprule
		\textbf{Symbol}&\textbf{Description}\\
		\midrule
		$G,V,E$& $G=(V,E)$, $V$ is a set of nodes, $E$ is a set of edges between nodes\\
		\midrule
		$n,m$& $n$ is the number of user nodes, $m$ is the number of edges\\
		\midrule
		$k$ & budget for the number of seeds\\
		\midrule
		$\sigma (\cdot) $ & the influence spread of a node (set)\\
		\midrule
		$p_e,p_{u,v}$& the propagation probability of edge $e$ or $(u,v)\in E$\\
		\midrule
		$\Delta(u|S)$& the marginal influence spread of node $u$ under active node set $S$\\
		\midrule
		$OPT$& the influence spread of the optimal seed set\\
		\midrule
		$\theta_v$& the threshold value of node $v$ being activated under LT model\\
		\midrule
		$\mathcal{X}_t$ & the active nodes within $G$ till step $t$\\
		\bottomrule
	\end{tabular}\vspace{-4ex}
\end{table}

\vspace{-2ex}\subsection{Social Network and Word-of-Mouth Effect}
A social network generally consists of a set of individuals and relationships among them. Depending on the nature of the network, the relationship can be friendship, subscription, follower, etc.
Formally, a social network is considered as a graph $G=(V,E)$, where $V$ is the set of nodes (\ie users) and $E$ is the set of (directed/undirected) edges in $G$ (\ie social links between users). In many cases, each edge has a weight that represents the strength of the relationship between a pair of users.

Given a social network, users can post their opinions, views, and comments, resulting in \textit{user generated content} (UGC). These information can be observed by their friends or followers, who may broadcast the information further by posting their own UGCs. Such process iterates through friends-of-friends/followers-of-followers and reaches a large population in the end. This phenomenon is referred to as ``word-of-mouth'' effect~\cite{DBLP:journals/tist/ZhangT11}. Specifically, users who post/broadcast an information are referred to as \textit{active} nodes (\ie adopters of the information) while the rest of the users are referred to as \textit{inactive} ones. This phenomenon may impact many real-world applications, \eg marketers can release free product samples for trial to a limited number of early adopters who may disseminate product-related information to other individuals through the effect; some governments may control public opinions in a social network by targeting a group of influential individuals who can potentially cause widespread dissemination of information.
In order to evaluate the scope of an information cascade due to the effect, it is vital to model under what condition(s) a user may post a UGC to broadcast the information (\eg retweet) that has been posted by her friends. This can be formally represented by an \textit{information diffusion} model (denoted by $M$) as follows.

\begin{definition}\label{def:difmod}
{\em Given $G=(V,E)$, for each node $v\in V$, let $\mathcal{N}(v)$ be the neighbors of $v$ and $\mathcal{X}_t$ be the activated nodes within $G$ till step $t$. An \textbf{information diffusion model}\footnote{For brevity, we shall refer to it as \textit{diffusion model} in the paper.} $M$ is used to evaluate whether node $v$ can be activated in step $t+1$ given that some of its neighbors are active at step $t$. $$[v\in \mathcal{X}_{t+1}]=M(\mathcal{N}(v)\cap\mathcal{X}_t)$$\/}
\end{definition}
Notably, $[\cdot]$ refers to a logistic evaluation and outputs 1 or 0 depending on whether `$\cdot$' is true or false. $M(\cdot)$ takes different forms and will be discussed in detail in the sequel.

\vspace{-2ex}\subsection{Definition of Influence Maximization Problem}
Given a social network $G$ and a piece of information to be disseminated, we use $S$ (\ie $\mathcal{X}_0$) to denote the initial users (\ie early adopters) who are going to initiate the information cascade. That is, all nodes in $S$ are \textit{active} before the information starts to spread. Nodes in $S$ can activate their friends, friend-of-friends, and so on based on a particular information diffusion model $M$. This process continues until there is no more nodes activated (\ie $\mathcal{X}_t=\mathcal{X}_\infty$). The final number of influenced nodes (\ie active nodes) at this step reflects the ability of $S$ to disseminate a piece of information. Formally, we introduce the notion of \textit{influence function} as follows to evaluate this ability.

\begin{definition}\label{def:inffunc} {\em
	Given $G=(V,E)$, a seed set $S$ (\ie early active nodes $\mathcal{X}_0$), the \textbf{influence function} of $S$, denoted as $\sigma(S)$, is the expected number of users influenced by $S$ where $\sigma(S)$ is a non-negative set function defined on any subset of users, \ie $\sigma:2^{V}\to \mathbb{R} \ge 0$. It can be formally computed as follows:
	$$\sigma(S)=E[|\mathcal{X}_\infty|]\mbox{ subject to }S=\mathcal{X}_0.$$
	In particular, for a node $v\notin S$, $\sigma(S\cup \{v\})-\sigma(S)$ is the \textbf{marginal influence} of $v$ with respect to $S$.\/}
\end{definition}

We are now ready to formally define the Influence Maximization (IM) problem.

\begin{definition}\label{def:im}
{\em Given $G=(V,E)$ and a positive integer $k$, the \textbf{influence maximization} problem aims to select a set $S^*$ of $k$ users from $V$ as the seed set to maximize the influence spread $\sigma(S^*)$, \ie $$S^*=arg max_{S\subseteq V \wedge |S| \le k}\sigma(S).$$\/}
\end{definition}

Notably, according to Definition~\ref{def:inffunc}, $\sigma(S)$ depends on $E[|\mathcal{X}_\infty|]$. Moreover, for any $t$, $\mathcal{X}_t$ depends on the specific diffusion model $M$ as well as $\mathcal{X}_{t-1}$, which is iteratively generated by $M$ and $\mathcal{X}_0$. \eat{Therefore, we have to first investigate the information diffusion model $M$ in the next section.}

\vspace{-1ex}\section{Influence Maximization Taxonomy}\label{sec2}
In this section, we present the framework of our survey by creating a taxonomy of the IM work in the literature. Since the goal of the IM problem is to maximize information diffusion, we first classify the diffusion models deployed in IM research. Next, we classify the large body of work in classical IM solutions. Finally, we categorize efforts that extend classical IM by incorporating various realistic issues associated with information diffusion.

Information diffusion models are fundamental to the IM problem as they guide the way information is propagated in a social network. In the literature, some models do not allow an \textit{active} node to become \textit{inactive} again whereas others do. Accordingly, existing diffusion models can be categorized into two types, namely \textit{non-progressive} and \textit{progressive} ones, respectively.

Based on the existing diffusion models, there is a large body of work on classical IM solutions. We broadly classify them w.r.t whether the quality of the influence spread is theoretically guaranteed or not. Specifically, we consider two categories of IM work, namely \textit{approximate} and \textit{heuristic} algorithms. The former provide theoretical guarantees whereas the latter ones do not. We further classify the approximate algorithms into \textit{simulation-based} and \textit{sampling-based}.\textit{ Simulation-based} methods leverage on Monte-Carlo simulations to estimate the influence spread $\sigma(\cdot)$ of a node (set). The \textit{sampling-based} algorithms aim to sample influence instance from a social graph $G$ to estimate the influence scope of a node (set). Generally, there are two types of samples that are widely adopted, namely \textit{snapshots} and \textit{reverse reachable (RR) sets}. The former estimates the influence spread of a node from the perspective of an ``influencer'', while the latter estimates from an ``influencee'' perspective. The heuristic algorithms can be further classified into \textit{local} and \textit{global} algorithms. The former category of work only utilize the local information of a network to guide the seeds selection whereas the latter ones consider the overall information of the entire social network.

\begin{figure}[t]
	\centering
	\includegraphics[width=0.55\textwidth,height=5cm]{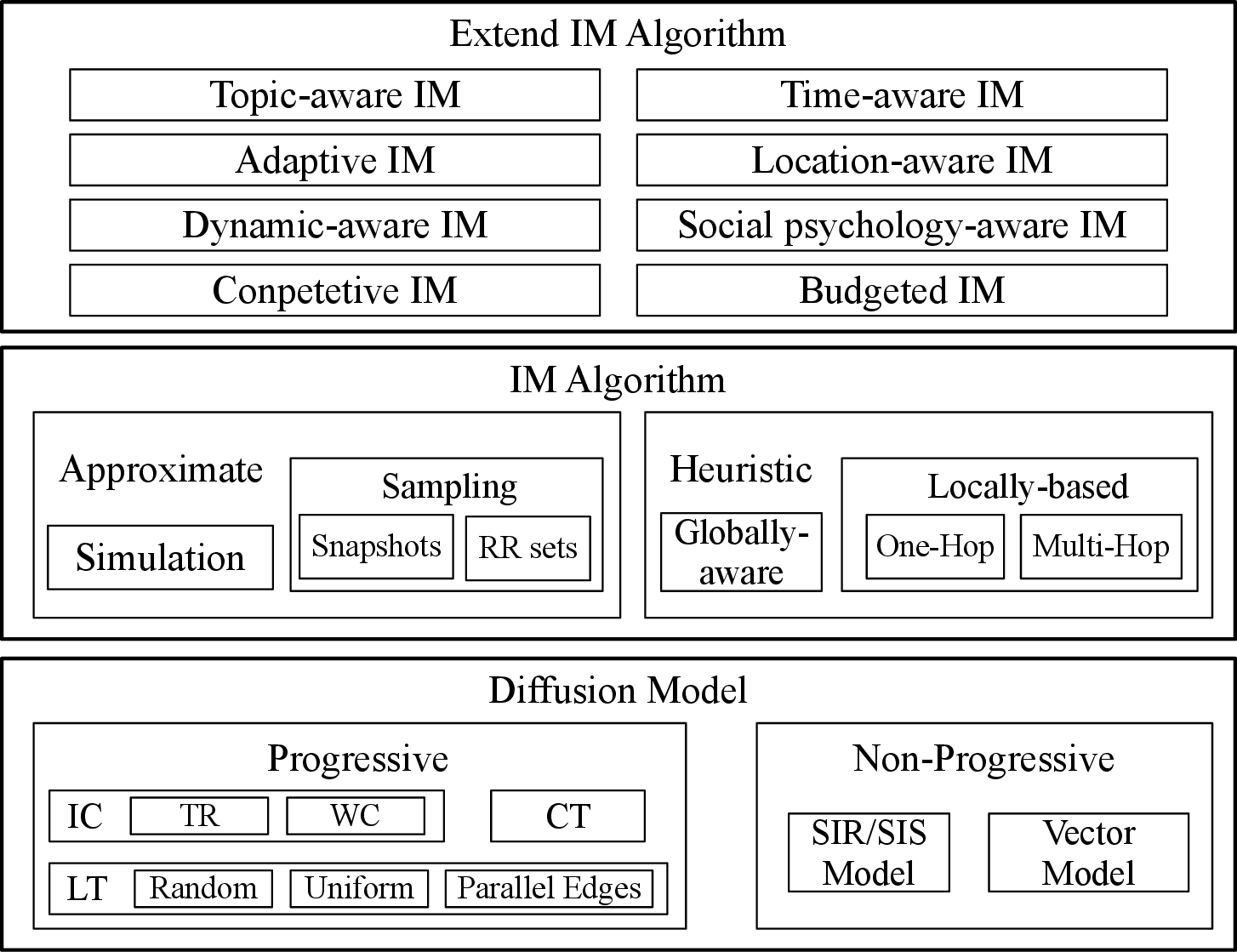}
	\vspace{-2ex}\caption{Framework of the survey.}
	\label{fig:fig_frame}\vspace{-4ex}
\end{figure}

There are many extensions of classical IM problem, which are classified as follows.

\begin{itemize} \itemsep=-.3ex
\item \textit{Topic-aware IM}. In a social network, users may often be interested in some topics such as music, sports, etc. In this extension category, several work aim to maximize the influence with respect to a given topic.
\item  \textit{Location-aware IM.} In location-based social networks (LBSN)~\cite{DBLP:conf/sigmod/LiCFTL14}, UGCs are often associated with geographical information. In this context, the goal of the IM problem is to maximize the influence within a given geographical region.
\item \textit{Social psychology-aware IM}. A limitation of classical IM techniques is that they strip off important human factors that impact social influence such as social psychology~\cite{citeulike:436278}. In this extension, the goal is to investigate the interplay between social psychology and the IM problem to devise more realistic solution.

\item \textit{Dynamic IM}. In reality, social networks are dynamic where new nodes and links are being added or deleted constantly. Maximizing the influence by considering this dynamic nature of a social network is another variation of classical IM.
\item \textit{Budgeted IM}. Given a budget and social network, different nodes will have different costs. The problem of budget IM is to select the seed node set in the budget to maximize the number of influenced nodes.
\item \textit{Time-aware IM}. Certain information in a social network can be time-sensitive (\eg a specific music concert). The goal of this body of work is to maximize the influence with respect to a given period or deadline.
\item \textit{Competitive IM}. In many real-world scenarios, several competing entities (\eg \textit{Samsung}, \textit{Apple}) may attempt to maximize the influence spread of the same information (\eg mobile phone) in the same social network synchronously or asynchronously. These competitors then form a competitive environment and the key goal of efforts in this extension is to maximize influence of an entity by taken the competition into account.
\item \textit{Adaptive IM}. In this recent flavor of IM, the node to initiate the spread of  information can be decided in runtime, \ie during the eventual dissemination. Hence, seed nodes can be incrementally selected based on the observed influence spread of the partial seeds selected in previous rounds.

\end{itemize}

Fig.~\ref{fig:fig_frame} depicts a pictorial overview of our framework based on which the rest of this survey is presented. A notable feature of our taxonomy is that, unlike~\cite{DBLP:journals/tkde/LiFWT18}, it clearly organizes and differentiates existing body of work in diffusion models, classical IM solutions, as well as extended IM techniques under one holistic framework.

\vspace{-1ex}\section{Diffusion Models}\label{sec4}
In the information spread process defined by Kemp \etal~\cite{DBLP:conf/kdd/KempeKT03}, each node $v$ in a social network $G$ can be \textit{active} (\ie an adopter of the information) or \textit{inactive} (\ie not yet an adopter). An \textit{inactive} node may change to \textit{active} status indicating that it has adopted the information. However, some models do not allow an \textit{active} node to become \textit{inactive} again. Depending on whether it is disallowed or allowed, existing diffusion models can be categorized into two types, namely \textit{progressive} and \textit{non-progressive} ones. In this section, we discuss them in turn.

\vspace{-2ex}\subsection{Progressive Model}
\textit{Progressive} diffusion models are popular among many existing influence maximization works that assume an \textit{active} node cannot be \textit{inactive} again during information spread. We list here the most representative models among this type.

\vspace{-1ex}\subsubsection{Independent Cascade Model}
Given a social network $G=(V,E)$, for each directed edge $(u,v)\in E$, $u$ is an in-neighbor of $v$ and $v$ is an out-neighbor of $u$. The \textit{Independent Cascade} (IC) model considers a user $v$ can be activated by each of its in-neighbors independently by associating an \textit{influence probability} $p_{u,v}$ to each edge $(u,v)$. Based on the influence probabilities and given a seed set $S$ at time step $0$, a diffusion instance of the IC model unfolds in discrete steps. Each active user $u$ in step $t$ will activate each of its out-neighbor $v$ that is inactive in step $t+1$ with probability $p_{u,v}$. The activation process can be considered as flipping a coin with head probability $p_{u,v}$. That is, if the result is head, then $v$ is activated; otherwise $v$ stays inactive. Note that $u$ has only one chance to activate its out-neighbors. After that, $u$ remains active and stops any further activations. The diffusion instance terminates when no inactive nodes can be activated. An example of influence propagation under IC model is depicted in Fig.~\ref{fig_IC}. Note that the influence probability $p_{u,v}$ on each edge significantly affects the result of information propagation. In particular, depending on how $p_{u,v}$ is defined, there are two types of IC, namely \textit{Weighted Cascade} model and \textit{Trivalency} model.
\eat{\begin{exmp}
	At $t=0$: $\mathcal{X}_0=\{v_1\}$; $t=1$: $\mathcal{X}_1=\{v_1,v_2\}$, eventual propagation=$\{v_1 \to v_2\}$; $t=2$: $\mathcal{X}_2=\{v_1,v_2,v_3,v_4\}$, eventual propagation=$\{v_2 \to v_3,v_2 \to v_4\}$; $t=3$: $\mathcal{X}_3=\{v_1,v_2,v_3,v_4,v_7\}$, eventual propagation=$\{v_3 \to v_7\}$; $t=4$: $\mathcal{X}_4=\{v_1,v_2,v_3,v_4,v_7,v_9\}$, eventual propagation=$\{v_7 \to v_9\}$.
\end{exmp}}
\begin{itemize} \itemsep=-.3ex
\item \btitle{Weighted Cascade Model (WC).} In this model, $p_{u,v}=\frac{1}{|In(v)|}$ where $In(v)$ is the set of in-neighbors of node $v$. In other words, each incoming neighbor influences $v$ with equal probability. As a result, it is easier to influence low-degree nodes than high-degree ones.

\item \btitle{Trivalency Model (TR).} In this model, the probability (or weight) on an edge is chosen randomly from a set of probabilities. For example, the weight may be chosen randomly from the set $\{0.001, 0.01, 0.1\}$.
\end{itemize}
\vspace{-2ex}\subsubsection{Linear Threshold Model}
\begin{figure*}[t]
	\centering
	\subfloat[IC] { \label{fig_IC}
		\includegraphics[width=0.47\columnwidth]{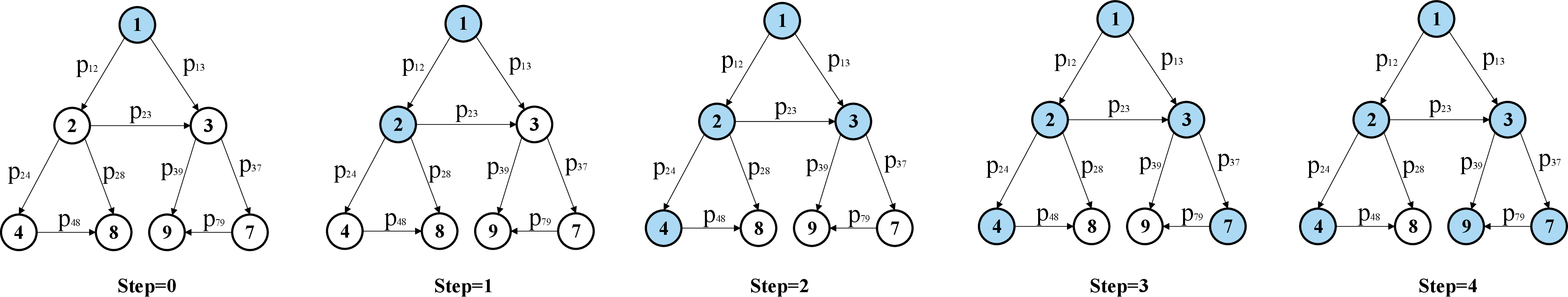}
	}\quad
	\subfloat[LT] { \label{fig_LT}
		\includegraphics[width=0.47\columnwidth]{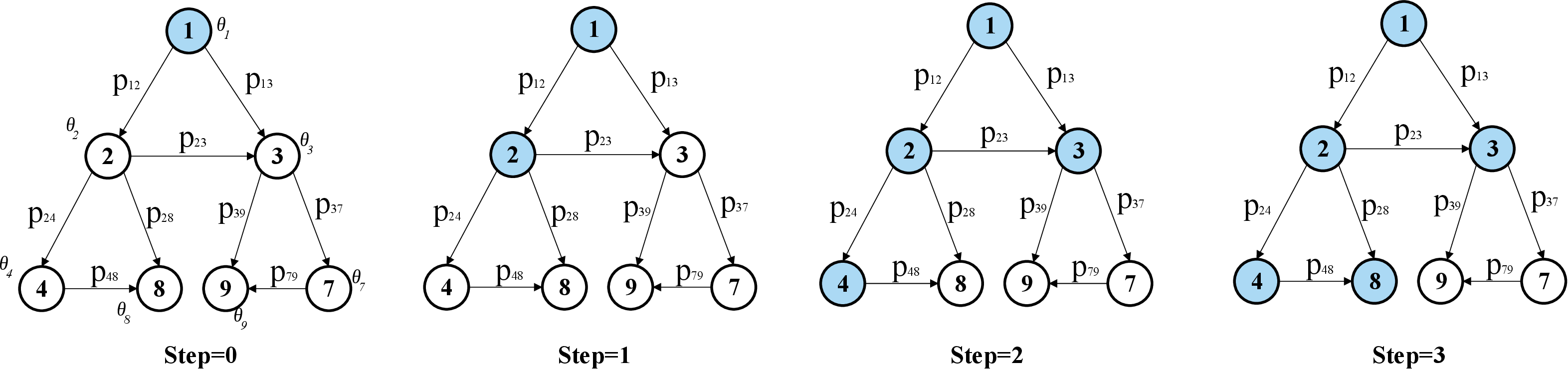}
	}\\
	\subfloat[Voter] { \label{fig_vt}
		\includegraphics[width=0.47\columnwidth]{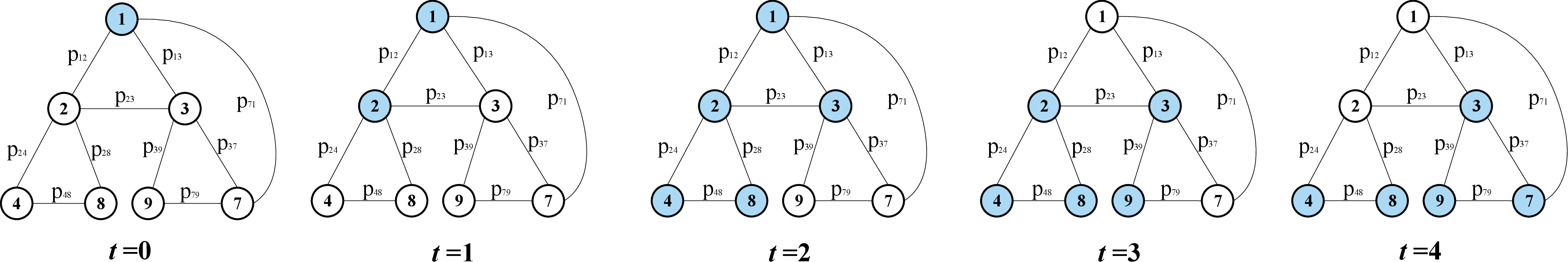}
	}\quad
	\subfloat[SIS] { \label{fig_SIS}
		\includegraphics[width=0.47\columnwidth]{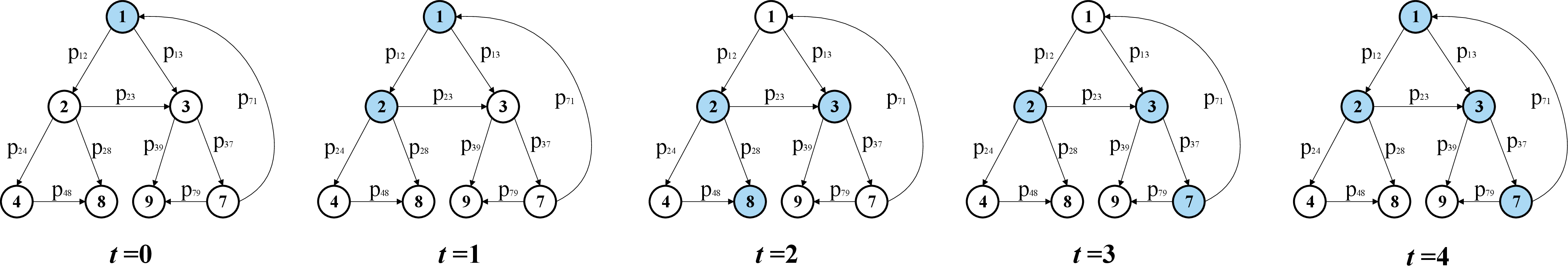}
	}
	\vspace{-3ex}\caption{Examples for different models (activated nodes are marked as blue)}
	\label{fig_models}
	\vspace{-3ex}
\end{figure*}

Given a social network $G=(V,E)$, in the \textit{Linear Threshold} (LT) model, each edge $(u,v)\in E$ is associated with a \textit{diffusion weight} $p_{u,v}$ defined by the function $p:E \to [0,1]$, subject to $\sum_{u}p_{u,v} \le 1, \forall v\in V$. If $(u,v) \notin E$, $p_{u,v}=0$. Meanwhile, each node $v$ chooses a threshold $\theta_{v}$ uniformly at random from the interval $[0,1]$, which represents the \textit{weighted threshold} of $v$'s in-neighbors that must be affected (\ie activated) to let $v$ become affected. Given the random choices of weighted thresholds of nodes and a seed set $S \subseteq V$, the LT model proceeds in discrete rounds $0,1,2,\ldots,$ as follows. Initially, at step $0$, nodes in $S$ are affected. Then at any step $t\ge 1$, a node $v$ becomes active if its threshold $\theta_{v}$ is surpassed by the total weights of its currently affected in-neighbors. That is, $\sum_{u\in\mathcal{X}_{t-1}}p_{u,v} \ge \theta_{v}$.
The process continues until no node can be activated. An example of influence propagation under the LT model is depicted in Fig.~\ref{fig_LT}.

\eat{\begin{exmp}
	At $t=0$: $\mathcal{X}_0=\{v_1\}$; $t=1$: $\mathcal{X}_1=\{v_1,v_2\}$, eventual propagation=$\{v_1 \to v_2\}$; $t=2$: $\mathcal{X}_2=\{v_1,v_2,v_3,v_4\}$, eventual propagation=$\{\{v_2,v_1\} \to v_3,v_2 \to v_4\}$; $t=3$: $\mathcal{X}_3=\{v_1,v_2,v_3,v_4,v_8\}$, eventual propagation=$\{\{v_2,v_4\} \to v_8\}$.
\end{exmp}}

The diffusion weight $p_{u,v}$ of an edge $(u,v)$ is generated by using one of the following strategies.

\begin{itemize}\itemsep=-.3ex
\item \btitle{Uniform.} The edge-weight on each edge $(u, v)$ is assigned to $p_{u,v}=\frac{1}{|In(v)|}$ . Observe that this is similar to the WC model.

\item \btitle{Random.} In this model, each edge is assigned a value selected uniformly at random in the range $[0,1]$. These values are then normalized to generate the edge weights so that weights of all incoming edges to a node $v$ sum up to $1$.

\item \btitle{Parallel edges.} In this model, each edge $(u, v)$ consists of a series of parallel links between $u$ and $v$, and has the weight
$
p_{u,v} =\frac{c(u,v)}{\sum_{\forall u^\prime \in In(v)}c(u^\prime,v)}
$, where $c(u,v)$ is the number of parallel edges from $u$ to $v$.\eat{ More simply, it is a generalization of the Uniform model for multi-graphs.}
\end{itemize}

\vspace{-2ex}\subsubsection{Other Progressive Models}
Lu \etal \cite{DBLP:journals/jco/LuZWKF12} proposed a Deterministic Linear Threshold (DLT) model, where the threshold can be regarded as the input of the model, instead of randomly generated. Compared with the LT model, when the threshold is fixed, the influence function is not submodular then. As a result, $1-1/e$ approximation ratio cannot be guaranteed in this setting. \cite{DBLP:journals/kbs/GursoyG18, DBLP:journals/mp/FischettiKLMR18, DBLP:journals/networks/GunnecRZ20} defined the new influence maximization problem under the DLT Model, and proposed several heuristic algorithms to solve this problem. 
%
The Continuous-Time (CT) model~\cite{DBLP:conf/icml/Gomez-RodriguezBS11} considers the likelihood of pairwise propagation between nodes as a continuous distribution of time. In particular, given a node $u$ activated at time $t_u$, the probability for $u$ to activate its neighbor $v$ at time $t_v$ ($t_v>t_u$), denoted as $p(t_v|t_u)$, is considered to be not only affected by the strength of connection between them but also by $t_v-t_u$. 
\vspace{-2ex}\subsection{Non-Progressive Model}
In contrast to progressive models (IC, LT, and CT) where active nodes cannot become inactive, in non-progressive models, an active node can be de-activated during the whole information spread process. The representative models in this category includes \textit{SIR/SIS} \cite{Kermack1927AG} and \textit{Voter} \cite{Clifford1973A,Holley1975Ergodic}.
%

\vspace{-1ex}\subsubsection{Voter Model}
In the original Voter model \cite{Clifford1973A,Holley1975Ergodic}, every node in a social network has two states, active and inactive, denoted as 0 and 1, respectively. Given an undirected social network $G=(V,E)$, which may contain self-loops, for a node $v\in V$, we denote by $\mathcal{N}(v)$ the set of neighbors of $v$ in $G$ (which includes $v$ itself since $G$ has loops). Starting from an arbitrary $0/1$ assignment to the nodes of $G$, at each step $t\ge 1$, each node picks its neighbor uniformly at random and adopts its opinion. Formally, starting form any initial node assignment $\mathcal{X}_0=S$, we inductively define
\begin{equation}\label{eq3}\footnotesize
[v\in\mathcal{X}_t]= 1 \mbox{ with probability } \frac{|\mathcal{N}(v)\cap\mathcal{X}_t|}{|\mathcal{N}(v)|} \mbox{ and }
0 \mbox{ otherwise.}
\end{equation}
The Voter model is a random process whose behavior depends on the initial assignment $\mathcal{X}_0$. For the IM problem under the Voter model, $v\in\mathcal{X}_t$ indicates that the node $v$ has adopted the information we wish to advertise. An example of influence propagation under the Voter model is shown in Fig.~\ref{fig_vt}. At $t=0$: $\mathcal{X}_0=\{v_1\}$; $t=1$: $\mathcal{X}_1=\{v_1,v_2\}$, diffusion instance=$\{\{v_1\} \to v_2\}$; $t=2$: $\mathcal{X}_2=\{v_1,v_2,v_3,v_4,v_8\}$, diffusion instance=$\{\{v_2,v_1\} \to v_3,\{v_2\} \to v_4,\{v_2\} \to v_8\}$; $t=3$: $\mathcal{X}_3=\{v_2,v_3,v_4,v_8,v_9\}$, diffusion instance=$\{\{v_3\}\to v_9,\{v_7\} \to v_1\}$; $t=4$: $\mathcal{X}_4=\{v_3,v_4,v_7,v_8,v_9\}$, diffusion instance=$\{\{v_1\} \to v_2,\{v_3,v_9\} \to v_7\}$. Hereby, \textit{diffusion instance} denotes the successful propagation attempts between consecutive steps.

\vspace{-1ex}\subsubsection{SIR/SIS Model}
The \emph{susceptible/infected/recovered model} (SIR) model and \textit{susceptible}/\textit{infected}/\linebreak\textit{susceptible} model (SIS) model are originally proposed in virus propagation study~\cite{Kermack1927AG}. In the SIR model, a node has three states, namely \textit{susceptible}, \textit{infected}, and \textit{recovered}, and only infected individuals can infect susceptible individuals, while recovered individuals can neither infect nor be infected by others. That is, an individual is never infected by a disease multiple times. In the SIS model~\cite{DBLP:journals/sigmod/GuilleHFZ13,DBLP:journals/siamrev/Hethcote00,DBLP:conf/sdm/LeskovecMFGH07}, once a node has been activated as susceptible at time $t$, it will remain susceptible or become infected at $t+1$. One infected node could become susceptible from infected status\eat{, considering that nodes have memories of the information they have been exposed to}. Fig.~\ref{fig_SIS} is an example of information propagation under the SIS model. Here, at $t=0$: $\mathcal{X}_0=\{v_1\}$; $t=1$: $\mathcal{X}_1=\{v_1,v_2\}$, diffusion instance=$\{v_1 \to v_2\}$, re-inactive nodes=$\emptyset$; $t=2$: $\mathcal{X}_2=\{v_2,v_3\,v_4\}$, diffusion instance=$\{v_2 \to v_4,v_2 \to v_3\}$, re-inactive nodes=$\{v_1\}$;  $t=3$: $\mathcal{X}_3=\{v_2,v_3,v_7\}$, diffusion instance=$\{v_3 \to v_7\}$, re-inactive nodes=$\{v_8\}$;$t=4$: $\mathcal{X}_4=\{v_1,v_2,v_3,v_7\}$, diffusion instance=$\{v_7 \to v_1\}$, re-inactive nodes=$\emptyset$.
%

\vspace{-1ex}\subsubsection{Heat Diffusion Model}
In social networks, early users (seed nodes) of products act as heat sources. Early users spread the influence to others as if a heat source continuously radiates heat, and the affected people will adopt the product. Ma \etal \cite{ DBLP:conf/cikm/MaYLK08} proposed the \emph{Heat Diffusion Model (HDM)}, which models the process of information diffusion as a physical phenomenon, heat diffusion. The HDM-based IM problem is defined as selecting a seed set and setting the initial heat for each node at the beginning of the thermal diffusion process. 
\cite{ DBLP:conf/cikm/MaYLK08, DBLP:journals/tist/ChenZPLL14} proposed heuristic algorithms to solve IM problem under HDM.
\vspace{-2ex}\section{Solutions for Classical IM}\label{sec5}


\begin{figure}[!t]\vspace{-4ex}
	\centering
	\includegraphics[width=0.7\textwidth,height=3.3cm]{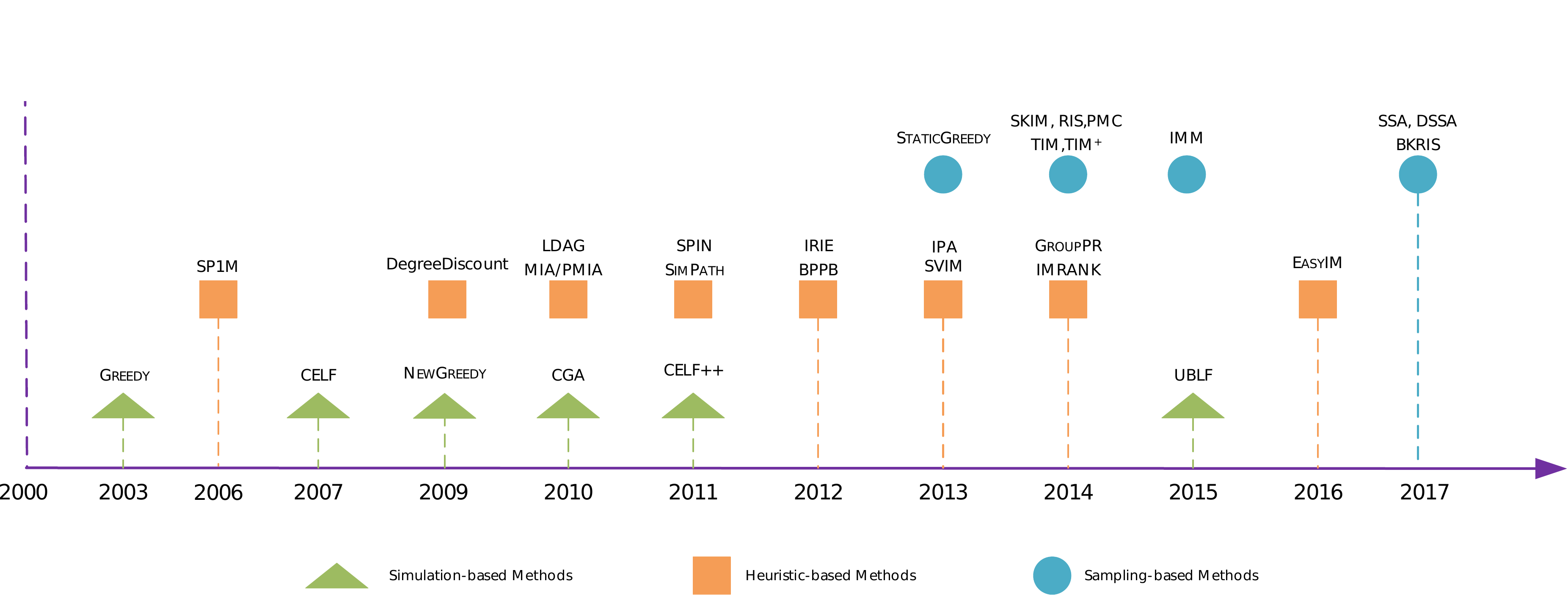}
	\vspace{-2ex}\caption{Representative works and milestones for classical IM solutions across all three categories}
	\label{fig_IMw}\vspace{-4ex}
\end{figure}

Based on the different diffusion models introduced in the preceding section, various research efforts have been proposed in the literature to address the classical IM problem as defined in Definition~\ref{def:im}. In this section, we shall conduct a thorough study of these efforts from three key dimensions: the diffusion model, the estimation method of the expected influence, and the selection of seed users. In this context, we shall also highlight their outstanding points and weaknesses, if any.

Since \cite{DBLP:conf/kdd/KempeKT03}, many works have emerged to accelerate the process for selecting seeds. Recall our discussion in Section~\ref{sec2}. Depending on whether the result quality can be guaranteed theoretically, these methods can be divided into two categories, namely approximate and heuristic algorithms. Based on the specific methods used to estimate the expected influence, approximate algorithms can be further classified into simulation-based and sampling-based. The key differences between these types of approaches can be summarized as follows.

\begin{itemize}\itemsep=-.3ex
\item \textit{Simulation-based algorithms.} This type of approaches estimates the influence spread of nodes by simulating the propagation of information.

\item \textit{Sampling-based algorithms.} These approaches aim to generate sufficient number of propagation samples and use them to estimate the influence spread of nodes. They typically take lesser time than simulation-based techniques.

\item \textit{Heuristic algorithms.} These algorithms sacrifice theoretical guarantees over results in order to gain in efficiency. The seeds selection process is guided by heuristic measures.
\end{itemize}

The developmental road-map of the aforementioned categories of representative IM efforts is shown in Fig.~\ref{fig_IMw}.

\vspace{-2ex}\subsection{Approximate Algorithms}\label{ssec:appro}
Kempe \etal \cite{DBLP:conf/kdd/KempeKT03} presented the first approximate algorithm based on hill-climbing strategy. Since then most of the existing IM algorithms follow the hill-climbing framework, which is illustrated in Algorithm \ref{alg:greedy}. The algorithm is initialized with an empty seed set $S$. It iteratively selects a node $u$ into $S$ if $u$ provides the maximum marginal gain to the influence function $\sigma(S)$ w.r.t. the partial solution $S$. The algorithm terminates when there are $k$ distinct nodes in $S$. The theoretical guarantee of this greedy framework depends on whether $\sigma(S)$ is a non-negative, monotone, submodular function. In particular, \cite{DBLP:conf/kdd/KempeKT03} proved that $\sigma(\cdot)$ is indeed so under some classical diffusion models such as IC and LT. Consequently, the greedy method leads to the result quality of $1-1/e-\varepsilon$ approximation to the optimal solution. Observe that the key challenge in Algorithm~\ref{alg:greedy} is to compute $\sigma(S)$ and $\sigma(S\cup \{u\})$. In \cite{DBLP:conf/kdd/KempeKT03}, Kempe \etal proved that evaluating $\sigma(\cdot)$ is a $\sharp P$-hard problem. Since then the majority of IM solutions present different methods to give unbiased estimation of $\sigma(\cdot)$ in polynomial time. These solutions can be grouped into two categories, namely simulation-based \cite{DBLP:conf/kdd/LeskovecKGFVG07,DBLP:conf/kdd/WangCSX10,DBLP:conf/www/GoyalLL11,DBLP:journals/tkde/ZhouZZG15} and sampling-based algorithms \cite{DBLP:conf/kdd/ChenWY09,DBLP:conf/cikm/ChengSHZC13,DBLP:conf/cikm/CohenDPW14,DBLP:conf/soda/BorgsBCL14,DBLP:conf/sigmod/TangXS14,DBLP:conf/sigmod/TangSX15,DBLP:conf/sigmod/NguyenTD16,DBLP:journals/tkde/WangZZLC17}.
We discuss them in turn.

\begin{algorithm}[t]
	\caption{\textsf{GREEDY}($k,\sigma$)}
	\label{alg:greedy}
	\scriptsize
	\begin{algorithmic}[1] 
		\REQUIRE ~~ $k$, a non-negative integer; $\sigma(\cdot)$, influence function.
		
		\ENSURE ~~ $S$, a seed set.
		\STATE $S \gets \emptyset$
		\FOR{$i=1,...,k$}
		\STATE $u^{*} \gets$ arg  $max_{u\in V\setminus S}(\sigma(S\cup \{u\})-\sigma(S))$
		\STATE $S \gets S \cup \{u^*\}$
		\ENDFOR
		\RETURN $S$
	\end{algorithmic}
\end{algorithm}

\vspace{-2ex}\subsubsection{Simulation-based Methods}
Simulation-based methods leverage on Monte-Carlo simulations to estimate the influence spread $\sigma(\cdot)$ of a node or a node set. In particular, it simulates the propagation process of the information under a specific diffusion model. In order to achieve an unbiased estimation of the influence spread, a large number of simulations must be undertaken, which is the key limitation for all these approaches.

The first work in this group can be found in \cite{DBLP:conf/kdd/KempeKT03}. Suppose the information propagates in a social network following the IC model and each edge $e$ is attached with an influence probability $p_e$. The process of MC simulation works as follows. Given a seed set $S$, we traverse the outgoing links from $S$ and simulate the node activation pseudo-randomly according to the IC model. That is, for an edge $(u,v)$, where $u$ is active but $v$ is not, we pseudo-randomly generate a value $\gamma$ from $[0,1]$. If $\gamma\ge p_{(u,v)}$, then $v$ is activated. This process continues until there is no new node activated. The above simulation only generates one instance from the possible world. In order to provide an unbiased estimation of the influence, this simulation is repeated many times. Kemp \etal~\cite{DBLP:conf/kdd/KempeKT03} suggested a lower bound for the simulation should be 10000 times and claimed that the result quality is comparable for 10000 simulations and 300000 simulations.
MC simulation gives an unbiased estimate of $\sigma(\cdot)$ but it is prohibitively expensive for large social networks. 

In order to improve the efficiency and scalability, Leskovec \etal proposed a \textit{lazy-forward} greedy algorithm \textsf{CELF} \cite{DBLP:conf/kdd/LeskovecKGFVG07}. It is motivated by an interesting phenomenon with respect to the marginal influence of each candidate seed. For ease of discussion, we denote by $S_i$ as the partial seed set after selecting $i$ seeds ($i\le k$), and let $\Delta(u|S_{i})$ be the marginal influence given the partial solution $S_i$, \ie $\Delta(u|S_{i})=\sigma(S_{i}\cup \{u\})-\sigma(S_{i})$. Under IC model, the marginal influence function is submodular, \ie
\begin{equation}\label{eq:marginalinf}\footnotesize
	\sigma(S_{i}\cup \{u\})-\sigma(S_{i}) \leq \sigma(S_{j}\cup \{u\})-\sigma(S_{j}) \mbox{ for any } S_{j}\subseteq S_{i}
\end{equation}
Therefore, $\Delta(u|S_{i})$ is upper bounded by $\Delta(u|S_{j})$ for any $j\le i$. Based on that, after selecting a new node, it is not necessary to recompute the marginal influence for all the rest nodes. Instead, the marginal influence for many of the rest nodes can be ignored without any simulation. For instance, suppose in the $i$-th iteration, the marginal influence of all non-seed nodes have been estimated and ranked in descending order as: $v_{i(1)},v_{i(2)},\ldots$ (\ie $\Delta(v_{i(1)}|S_{i-1})\ge\Delta(v_{i(2)}|S_{i-1})\ge\Delta(v_{i(3)}|S_{i-1})\ge\ldots$). Hence $v_{i(1)}$ is selected into $S$, leading to $S_i$, and the algorithm continues to the $(i+1)$-th iteration. At this point, the marginal influence of $v_{i(2)}$ is re-estimated first, resulting in $\Delta(v_{i(2)}|S_{i})$. Afterwards, instead of re-estimating the marginal influences of $v_{i(2)},\ldots$ as~\cite{DBLP:conf/kdd/LeskovecKGFVG07}, \textsf{CELF} compares $\Delta(v_{i(2)}|S_{i})$ against $\Delta(v_{i(3)}|S_{i-1})$ first. If $\Delta(v_{i(2)}|S_{i})$ is larger, then $v_{i(2)}$ should be selected in this iteration and the marginal influence of all the rest nodes do not need to be re-estimated. This is because according to Eq.~\ref{eq:marginalinf}, $\Delta(\cdot|S_{i})$ for any node in $\{v_{i(3)},v_{i(4)},\ldots\}$ is upper bounded by $\Delta(v_{i(3)}|S_{i-1})$. In this manner, a lot of re-estimation over non-seed nodes can be avoided. Notably, \textsf{CELF} visits all nodes in $V\verb|\|S_{i-1}$ in descending order of their upper bounds of $\Delta(\cdot|S_{j})$, and computes $\Delta(u|S_{i-1})$ using MC simulations. Observe that this strategy does not improve the worst-case time complexity. However, this strategy brings up to 700 times improvement in practical performance compared to Algorithm~\ref{alg:greedy} in~\cite{DBLP:conf/kdd/KempeKT03}.


To further improve on the efficiency in estimating the influence spread via MC simulation, \textsf{CELF++} \cite{DBLP:conf/www/GoyalLL11} proposed a more effective pruning strategy based on \textsf{CELF}. \textsf{CELF++} maintains a tuple $\langle u.mg1, u.pre\_best, u.mg2, u.flag\rangle$ for each node $u$, where $u.mg1$ denotes the marginal influence of $u$ with respect to the current seed set $S_i$ and $u.pre\_best$ refers to the node with the maximum marginal gain in the current iteration, $u.mg2=\Delta(u|S_i\cup \{u.pre\_best\})$, and $u.flag$ is the iteration number when $u.mg1$ was last updated. Given the tuple for $u$, whenever $u.pre\_best$ is selected as the next seed in the current iteration, it doesn't need to re-estimate the marginal gain of $u$ in the next iteration. In this way, \textsf{CELF++} further filters a series of computational steps but maintains the same time complexity. A recent benchmarking study \cite{DBLP:conf/sigmod/AroraGR17}, however, demonstrated that \textsf{CELF++} does not exhibit significant advantage in running time compared to \textsf{CELF}.

Instead of estimating the approximate influence spread of nodes and ordering them, Zhou \etal \cite{DBLP:journals/tkde/ZhouZZG15} proposed a more efficient method called \textsf{UBLF}. During each iteration, nodes are ordered by the upper bound of their expected influence. Let $\sigma^\prime$ denote a vector where each element is $\sigma(u)$ for $u\in V$. \textsf{UBLF} presents an upper bound estimation method as $\sigma^\prime \le \sum_{i=1}^{n}PP^i \cdot \mathbf{1}$,
where $PP$ is the \textit{propagation probability matrix} associated with the social network and $\mathbf{1}$ denotes a column vector with each entry being $1$. Specifically, each entry $PP_{ij}$ refers to the propagation probability from node $v_i$ to $v_j$. \eat{The process of calculating $PP^i$ is fast and as $PP$ is sparse.} Afterwards, \textsf{UBLF} obtains the node $u$ which has the maximum upper bound of $\sigma$ and estimates its marginal influence $\sigma(u)$ using Monte-Carlo simulation. \eat{If $\sigma(u)$ is already larger than the upper bound of other nodes, it is not necessary to estimate their marginal influences in this iteration.} Empirically, \textsf{UBLF} reduces more than 95\% Monte-Carlo simulations of \textsf{CELF}~\cite{DBLP:conf/kdd/LeskovecKGFVG07} and achieves about 2-10 times speedup at $k=50$ when the propagation probabilities is less than $0.01$. However, its running time increases exponentially with respect to the propagation probability whereas \textsf{CELF} exhibits linear increase.\eat{ Therefore, empirical results justify that \textsf{UBLF} is inferior to \textsf{StaticGreedy} when the propagation probability is larger than 0.08.}

\textsf{CGA} algorithm~\cite{DBLP:conf/kdd/WangCSX10} proposed an effective method to get influential user nodes from a new perspective. It is motivated by the following property that has been widely studied in various research~\cite{Girvan7821,DBLP:conf/www/DongSXW09}: ``\textit{a community is a densely connected subset of nodes that are only sparely linked to the remaining network, and individuals in a community will influence each other in the form of `word-of-mouth'}''. Specifically, in the first step, \textsf{CGA} divides the target social network into several partitions according to the relationships between nodes, forming communities. Next, it utilizes the influence of each node within its community to determine which nodes are selected as seeds. Different from the aforementioned simulation-based algorithms, \textsf{CGA} only performs MC simulations on local subgraphs. It produces a $(1-\frac{1}{e^{1+\delta \rho}})$ approximate solution for IM, where $\rho$ is a pre-defined threshold determining the tightness of the extracted community, and $\delta$ is a parameter controlling the accuracy loss of influence estimation caused by graph partitioning.

\vspace{-1ex}\subsubsection{Sampling-based Methods}\label{ssec512}
Different from the simulation-based algorithms, the key idea of sampling-based algorithms is to sample \textit{influence instances} from $G$ to estimate the influence scope of a node (set). Intuitively, given that the influence propagation is viewed as a stochastic process within the target graph, all possible influence instances construct a possible world. Samples from this possible world can be utilized to estimate the expected influence of a node set. As mentioned earlier, there are two types of samples that are widely adopted in IM research, namely \textit{snapshots} and \textit{RR sets}. Both strategies share a common first step, flipping a coin on each edge $e$ with $p_e$ (sampling-based approaches mainly work under the IC model). The difference between them is manifested in the second step. The snapshot strategy typically estimates the influence spread of a node from the perspective of ``influencers'' whereas the RR sets-based strategy ``reverses'' $G$ to consider the perspective of ``influencees''  in order to estimate the influence of nodes.

\noindent\textbf{Snapshots-based Sampling.}
Chen \etal \cite{DBLP:conf/kdd/ChenWY09} introduced the sampling approach in estimating the influence of a node (set). It is motivated by the following fact under IC: whether $u$ can activate $v$ depends only on $p_{u,v}$, which can be considered as flipping a coin of bias $p_{u,v}$. They proposed the \textsf{NewGreedy} algorithm. It randomly generates a sample from the possible world of all influence instances given a network $G$, where each sample is referred to as a \textit{snapshot} defined as follows.

\vspace{-1ex}\begin{definition}[Snapshot]
	{\em Given a graph $G=(V,E)$, we flip all coins a priori and remove each edge $(u,v)\in E$ with probability $1-p_{u,v}$, resulting in a subgraph $G^\prime=(V^\prime,E^\prime)$, which is called a \textbf{snapshot}.}
\vspace{-2ex}\end{definition}
%

In fact, $G^\prime$ can be viewed as a sample from the possible world of all influence instances. The influence scope of any node in $G^\prime$ can be easily evaluated by traversing $G^\prime$. Notably, this only contributes to the influence with respect to a particular instance in the possible world. The expectation of the influence for a node (set) has to take into account the influence scope over all instances within the possible world. In order to provide an unbiased estimation for the expectation of influence, \textsf{NewGreedy} iteratively generates $R$ (usually set to $20000$) instances and the average influence scope over all these influence instances is used as the estimation for the expected influence scope for a node (set). \eat{For instance, Fig.~\ref{fig.sampling} illustrates an example of generating three influence instances under the IC model.} Chen \etal further proposed \textsf{MixGreedy} algorithm which combines \textsf{NewGreedy} and the lazy-update strategy in~\cite{DBLP:conf/kdd/LeskovecKGFVG07}. 

Cheng \etal \cite{DBLP:conf/cikm/ChengSHZC13} proposed a dynamic update strategy called \textsf{StaticGreedyDU} that exploits the advantage of static snapshots and calculates the marginal gain in an efficient incremental manner. Specifically, when a node $v^*$ is selected as a seed, it directly discounts the marginal gain of other nodes by the marginal gain shared by these nodes and $v^*$. According to~\cite{DBLP:conf/cikm/ChengSHZC13}, \textsf{StaticGreedyDU} empirically reduces the computational cost by two orders of magnitude without loss of accuracy.

Pruned Monte-Carlo method (\textsf{PMC}) proposed in \cite{DBLP:conf/aaai/OhsakaAYK14} also aims to enhance the scalability of IM techniques. \eat{Specifically, \textsf{PMC} consists of the following steps:
(1) generate a random influence instance from graph $G$, (2) construct vertex-weighted directed acyclic graphs (DAGs) from the influence instance, (3) estimate the value of marginal influence by averaging the total weight of vertices reachable from a single vertex in each DAG, (4) select a seed according to the greedy strategy, and (5) update the DAGs. The key points in this method are how to generate DAGs and how to update them. To this end,} \textsf{PMC} computes strongly connected components (SCCs) from a random instance $G_i^\prime=(V,E_i^\prime)$ such that it can obtain a SCC containing $v\in V$ (denoted as $comp_i[v]$) as well as the number of vertices in $comp_i[v]$, denoted as $weight_i[v]$. Afterwards, the $i$-th vertex-weighted DAG $G_i=(V_i,E_i)$ could be obtained, where $V_i=\{comp_i[v]|v\in V\}$ and $E_i=\{( comp[u]),comp[v])|(u,v)\in E_i^\prime\}$. A breadth-first search (BFS) is then conducted over $comp_i[v]$ to return the marginal gain of $v$ in $G_i$, which is $\sigma_{G_i}(S\cup\{u\})-\sigma_{G_i}(S)$. An estimated value of marginal influence $\sigma(S\cup\{u\})-\sigma(S)$ is obtained by averaging the marginal gain of $v$ in each DAG $G_i$. After selecting a node $v$ with the maximum average gain, it updates the DAGs by removing vertices reachable from $comp_i[v]$ in $G_i$. \eat{Since BFSs still need to be conducted approximately for $knR$ times, a pair of speed-up strategies are introduced in the work. One works by pruning the number of vertices visited during BFS, while another aims to avoid redundant marginal gain recomputation.} Empirical study demonstrates that \textsf{PMC} exhibits similar running time with \textsf{IRIE}~\cite{DBLP:conf/icdm/JungHC12}, one of the state-of-the-art heuristic approach at that time (\ie 2014).

Unfortunately, during the same year when \textsf{PMC} was proposed~\cite{DBLP:conf/aaai/OhsakaAYK14}, another snapshot approach called \text{SKIM}~\cite{DBLP:conf/cikm/CohenDPW14} showed superior efficiency to \textsf{IRIE} while providing theoretical guarantee over the results quality. \eat{It is a \textit{sketch-based} influence maximization method using a bottom-$k$ minhash sketch~\cite{DBLP:conf/podc/CohenK07} that provides theoretical guarantee over the results quality while maintaining high efficiency.} Specifically, \textsf{SKIM} first samples a series of propagation instances $G^i$ like \textsf{MixGreedy}~\cite{DBLP:conf/kdd/ChenWY09}. Afterwards, it randomly assigns each node $v$ that is reachable from $u$ in $G^i$ a rank value sampled from uniform distribution within $[0,1]$. Next, the bottom-$k$ ranked nodes form a bottom-$k$ minhash sketch~\cite{DBLP:conf/podc/CohenK07} of $u$, denoted as $X_u$. The maximum rank values among $X_u$ are then used to estimate the influence of node $u$. Experiments show that in contrast to other snapshot-based sampling approaches, \textsf{SKIM} scales to graphs with billions of edges. Besides, it demonstrates comparable performance with a representative RR-set-based sampling method called \textsf{TIM}$^+$~\cite{DBLP:conf/sigmod/TangXS14}, \ie the same result quality but different efficiency (relative difference of running time between \textsf{SKIM} and \textsf{TIM}$^+$ varies from 0.24 to 3.19 over 11 datasets as reported in~\cite{DBLP:conf/cikm/CohenDPW14}).

\noindent\textbf{RR sets-based Sampling.} Borgs \etal~\cite{DBLP:conf/soda/BorgsBCL14} proposed a drastically different method, namely \textit{Reverse Influence Sampling} (\textsf{RIS}), to estimate the influence spread of a node from a mathematical point of view. In particular, in this method the number of samples from $G$ is typically lesser compared to \textsf{\textsf{StaticGreedy}}. \textsf{RIS} introduces a new sampling concept called \textit{Reverse Reachable Set} (RR set) to estimate the influence of given nodes.

\eat{In \cite{DBLP:conf/kdd/KempeKT03}, Kempe et al. pointed that the $(1-1/e-\varepsilon)$ approximation factor is nearly optimal, as no polytime algorithm achieves approximation with $(1-1/e+\varepsilon)$ for any $\varepsilon\ge 0$ unless $P=NP$, and as RIS is randomized, it succeeds with probability $1-1/n$; moreover, failure is detectable, so this success probability can be amplified through repetition.}

\vspace{-1ex}\begin{definition}[Reverse Reachable Set]
	{\em Let $v$ be a node in $V$ and $G^\prime$ be a graph by removing each edge $e$ in $G$ with probability $1-p_e$. The \textbf{Reverse Reachable Set} (\textbf{RR set}) for $v$ in $G^\prime$ is the set of nodes in $G^\prime$ that can reach $v$. That is, for each node $u$ in the RR set, there is a directed path from $u$ to $v$ in $G^\prime$.}
\vspace{-1ex}\end{definition}
\begin{definition}[Random RR Set]
	{\em Let $\mathcal{G}$ be the distribution of $G^\prime$ induced by the randomness of edge removals from $G$. A \textbf{Random RR set} is an RR set generated on an instance of $G^\prime$ randomly sampled from $\mathcal{G}$ for a node selected uniformly at random from $G^\prime$.}
\vspace{-1ex}\end{definition}

\begin{figure}[t]
	\centering
	\includegraphics[width=0.55\textwidth,height=3.4cm]{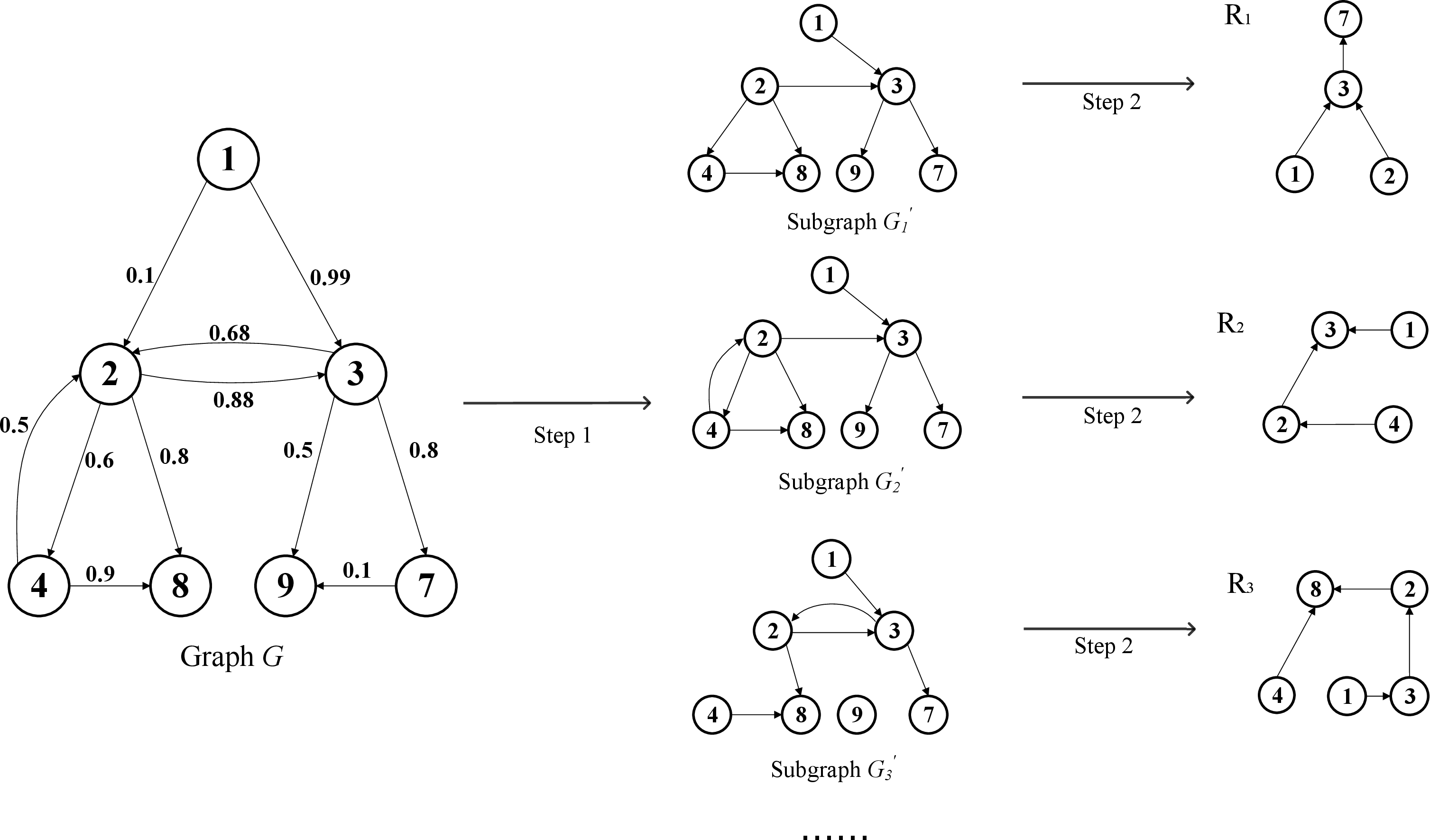}
	\vspace{-3ex}\caption{The process of generating RR sets. Step 1: sampling from the graph $G$, removing every edge $e$ with probability $1-p_e$; Step 2: select a node uniformly at random, and perform reverse BFS/DFS from this node.}
	\label{rr set}\vspace{-3ex}
\end{figure}

\textsf{RIS} constructs a sample polling in which RR sets are generated randomly as follows: select a node $v$ uniformly at random and determine the set of nodes that would have influenced $v$. This process terminates when a sufficiently large set $\mathcal R$ of random RR sets have been generated, and the size of $\mathcal R$ is set to $144(m+n)\varepsilon^{-3}\log n$ in RIS algorithm. Fig.~\ref{rr set} illustrates an example of RR set polling with three instances. In the RR set polling, the probability that a node $u$ appears in RR sets is proportional to $\sigma(u)$, \cite{DBLP:conf/soda/BorgsBCL14} further justified that this probability can be estimated accurately with relatively few repetitions of the polling process, \ie the more a node covers RR sets, the larger its influence spread will be. Based on the RR set polling, an \textit{index} for all nodes in $G$ could be constructed, where each node $u$ maintains a set $I_{u}$ containing the sequence numbers of RR sets covered by it. During each iteration of seeds selection, the node with the largest $I_u$ should be adopted. The key contribution of \textsf{RIS} algorithm is that it reduces the time complexity to $O(k\varepsilon^{-3}(n+m)\log^2 n)$.

In order to get a tighter lower bound for the number of RR sets, referred to as $\theta$, \textsf{TIM} \cite{DBLP:conf/sigmod/TangXS14} presents the first approach to minimize the number of RR sets required for unbiased estimation of influence with respect to a node (set). In particular, \textsf{TIM} defines a parameter
$
\lambda=(8+2\varepsilon)n\cdot(\mathit{l}\log n+\log\dbinom{n}{k}+\log 2 )\cdot \varepsilon^{-2}
$	
and proved that when $\theta\ge \lambda/OPT$, the greedy algorithm built upon RR sets returns a $(1-1/e-\varepsilon)$-approximation with $1-1/n^{l}$ probability, where $l$ (usually set to $1$~\cite{DBLP:conf/soda/BorgsBCL14,DBLP:conf/sigmod/TangXS14}) is a user-specified factor that controls the trade-off between result quality and efficiency. Moreover, the expected time complexity of \textsf{TIM} is $O(\theta \cdot EPT)$, where $EPT$ is the expected number of coin tosses required to generate an RR set for a randomly selected node in $G$. Based on this, \textsf{TIM} also reveals how to obtain the value of $\theta$ that makes $\theta \cdot EPT$ reasonably small, while ensuring $\theta\ge \lambda/OPT$.\eat{ Additionally, \textsf{TIM} supports the triggering model \cite{DBLP:conf/kdd/KempeKT03}, which is a general cascade model where IC model can be viewed as a special case.} \textsf{TIM} has a time complexity of $O((k+\mathit{l})(m+n)\log n\cdot \varepsilon^{-2})$. Empirically, it is superior to \textsf{RIS} while offering a strong theoretical guarantee. Tang \etal~\cite{DBLP:conf/sigmod/TangXS14} also presented $TIM^{+}$, which further improves the practical performance of \textsf{TIM} (without affecting its asymptotic guarantee) by proposing a much tighter lower-bound of $OPT$. \textsf{TIM}$^{+}$ has the same theoretical guarantee with respect to the approximation ratio but better empirical performance.

Since \textsf{TIM}, there is a series of efforts to improve efficiency and robustness of the RR set-based sampling strategy. The authors of \textsf{TIM} proposed another advanced approach based on \textsf{TIM}, namely \textsf{IMM} \cite{DBLP:conf/sigmod/TangSX15}. Compared to \textsf{TIM}/\textsf{TIM}$^{+}$, \textsf{IMM} improves the efficiency significantly by providing a tighter lower bound for the number of RR sets required to be generated. Empirical study in~\cite{DBLP:conf/sigmod/TangSX15} demonstrate that \textsf{IMM} is orders of magnitude faster than \textsf{TIM}/\textsf{TIM}$^+$ under IC and LT models. Notably, although \textsf{IMM} is claimed in \cite{DBLP:conf/sigmod/TangSX15} to achieve the same result approximation ratio as \textsf{TIM}, \ie $(1-1/e-\varepsilon)$-approximation with $1-1/n^{l}$ probability, a recent study~\cite{chen2018issue} pointed out that the `$1-1/n^{l}$ probability' can only be satisfied when $\theta$ is fixed. However, the sampling procedure in the original \textsf{IMM} algorithm will return a random $\theta$ value, such that the `$1-1/n^{l}$ probability' cannot be satisfied. Moreover, this issue has been acknowledged by the original author of \textsf{IMM}~\cite{chen2018issue}. To fix the problem, Chen \etal presented a pair of modifications over the original \textsf{IMM}, which incur some extra (less than 100\% empirically) computational overhead~\cite{chen2018issue}.

A different comparison result between \textsf{IMM} and \textsf{TIM}$^+$ has been reported in an experimental study~\cite{DBLP:conf/sigmod/AroraGR17}. This study showed that both of them have similar running time, and do not scale in terms of memory-consumption under IC model. \textsf{TIM}$^+$ is even faster than \textsf{IMM} in LT model\footnote{\scriptsize Lu \etal~\cite{DBLP:journals/corr/LuX0HL17} have refuted these results, which is counter-refuted by the original authors.}.\eat{ However, a later study~\cite{DBLP:journals/corr/LuX0HL17} presented a series of refutations towards~\cite{DBLP:conf/sigmod/AroraGR17}. Firstly, the memory-consumption test conducted in~\cite{DBLP:conf/sigmod/AroraGR17}, where $\varepsilon$ is fixed to 0.05 for both \textsf{IMM} and \textsf{TIM}$^+$, are unfair. It requires both algorithms to generate extremely accurate results, which definitely increases the computation overhead. In contrast, the other algorithms compared in the same test are allowed to produce less accurate results. Secondly, the claim that ``\textsf{TIM}$^+$ is even faster than \textsf{IMM} in LT model'' is invalid. In~\cite{DBLP:conf/sigmod/AroraGR17}, they evaluate both algorithms' running time to achieve similar empirical accuracy and find that \textsf{TIM}$^+$ spend less time. However, during the comparison, $\varepsilon$ is set to different values in both methods (0.1 in \textsf{TIM}$^+$ and 0.05 in \textsf{IMM}). Recall that $\varepsilon$ controls \textit{theoretical worst-case accuracy} of RR set-based solutions, it means that \textsf{IMM} will provide higher theoretical guarantee over the result accuracy in this setting. Given a higher $\varepsilon$, a larger number of RR sets have to be generated for both algorithms. Although they may share similar empirical results in some scenarios under such setting, the experimental results are meaningless as the comparison are conducted based on different theoretical accuracy guarantees.}
\eat{ Moreover, \textsf{IMM} performs well under CTIC model introduced in~\cite{DBLP:conf/icml/2011}.}

Hung \etal proposed a pair of novel algorithms \textsf{SSA} and \textsf{D-SSA} \cite{DBLP:conf/sigmod/NguyenTD16} based on \textsf{IMM} \cite{DBLP:conf/sigmod/TangSX15} to further improve the efficiency. \textsf{SSA} first generates $\Lambda$ RR sets denoted as $\mathcal{R}_1$ to get a seed set $S_k$. Afterwards, it presents two early-stop criterion and checks whether they are satisfied. One stopping condition is the number of $S_k$ covering $\mathcal{R}_1$ needs to be larger than $\Lambda$. If this condition holds, it will test against the second criteria, otherwise double the number of the RR sets and get the new seed set $S_k$. The second criteria is based on the \textit{zero-one estimation theorem} introduced in~\cite{DBLP:journals/siamcomp/DagumKLR00} and works as follows. \textsf{SSA} adopts another set of RR sets called $\mathcal{R}_2$ with the same size as $\mathcal{R}_1$ to estimate the expected influence spread $\hat{I}_{S_k}$ of $S_k$. Whenever $\hat{I}_{S_k}$ exceeds a pre-defined threshold, it returns the seed set $S_k$. In \cite{DBLP:conf/sigmod/NguyenTD16}, an improved method of \textsf{SSA}, namely \textsf{D-SSA} is also proposed. In \textsf{SSA} there are three internal parameters, namely $\varepsilon_1, \varepsilon_2, \varepsilon_3$, all of which influence the approximation ratio of the algorithm and determine the number of random RR sets required during influence estimation. In \textsf{SSA}, these parameters are fixed as $\varepsilon_1=\frac{1}{6}\varepsilon$, $\varepsilon_2=\frac{1}{2}\varepsilon$, $\varepsilon_3=\frac{1}{4(1-1/e)}\varepsilon$. According to \textsf{D-SSA}, these internal parameters are not fixed and can vary during iterations, such that the number of random RR sets generated can be reduced. Both \textsf{SSA} and \textsf{D-SSA} are claimed to provide $(1-1/e-\varepsilon)$ approximation ratio according to the analysis in~\cite{DBLP:conf/sigmod/NguyenTD16}.

However, a following-up study~\cite{DBLP:journals/pvldb/HuangWBXL17} revisited \textsf{SSA}/\textsf{D-SSA} and pointed out two issues: (i) the reported accuracy and efficiency results for \textsf{SSA}/\textsf{D-SSA} are inaccurate; (ii) there exist important gaps in the proof of \textsf{D-SSA}'s efficiency in~\cite{DBLP:conf/sigmod/NguyenTD16}. In particular, the empirical study in~\cite{DBLP:journals/pvldb/HuangWBXL17} showed that the implementation for some common steps in \textsf{SSA}/\textsf{D-SSA} and \textsf{IMM} are not optimized to the same level. \textsf{SSA}/\textsf{D-SSA} implementation adopted a faster random number generator and more efficient method to produce random RR sets under the LT model. Besides, they also adopted a more optimized implementation of the greedy algorithm to find seed sets from random RR sets. In spite of this, the overall efficiency of \textsf{IMM} is still inferior to that of \textsf{SSA} and \textsf{D-SSA}. \textsf{IMM} only outperforms these techniques when $k<20$. Another work~\cite{DBLP:conf/csonet/NguyenDT18} immediately \textit{revisited} the revisit in~\cite{DBLP:journals/pvldb/HuangWBXL17} and affirmed the issue (i) but not the issue (ii). They provided detailed explanations to justify that the theoretical study of the efficiency for \textsf{D-SSA} in~\cite{DBLP:conf/sigmod/NguyenTD16} is correct.

Instead of limiting the number of RR sets generated, \textsf{BKRIS}~\cite{DBLP:journals/tkde/WangZZLC17} pays attention to speeding up the seeds selection phase. The main idea of \textsf{BKRIS} is to first sort the RR sets in descending order based on an oracle in which a total order is assigned to all the samples. Such an order demonstrates a good property that if a node covering more samples is ranked in the front, it is more likely to have larger influence. The method to order the samples is achieved with the help of the bottom-$k$ sketch. Specifically, \textsf{BKRIS} first defines a threshold $bk$, generates a reasonable size of RR sets and order them. Whenever the number of samples covered by a node reaches $bk$, \textsf{BKRIS} returns it as the first seed. Moreover, for the samples size $\theta$,  \textsf{BKRIS} utilizes a heuristic method to estimate a lower bound of \textit{OPT}. Experiments on several large datasets show that it achieves up to two orders of magnitude speedup compared to \textsf{IMM} on both IC and LT models. It also achieves a $(1-1/e-\varepsilon-\varepsilon^\prime)$-approximation ratio when the $bk$ value for the bottom-$k$ sketch is $O(k^2 \varepsilon'^{-2}\log{n^{2+\log_nk}})$ where $\varepsilon$ is the error for estimating the influence function using RR-sketch and $\varepsilon^\prime$ is the error for estimating each node's coverage by the bottom-$k$ sketch.

Recently, \cite{DBLP:conf/icdm/NguyenNPD17, DBLP:conf/sigmod/TangTXY18, DBLP:journals/ton/NguyenTD17, DBLP:conf/sigmod/Guo0WC20} proposed improved algorithms for RIS, either from the approximation rate or running time. Nguyen \etal \cite{DBLP:conf/icdm/NguyenNPD17} proposed an improved sampling algorithm of the RIS, called SKIS, which deletes singular samples (samples of size one) in the sketch and only returns non-singular samples in the network. As a result, when the influence probability is small, the estimation quality can be improved and the sketch generation can be accelerated. Tang \etal \cite{DBLP:conf/sigmod/TangTXY18} proposed the OPIM-C algorithm, which starts from generating a small number of RR sets, iteratively adding new ones, and terminates immediately after meeting the approximate ratio. In this way, the running time is saved. Nguyen \etal \cite{DBLP:journals/ton/NguyenTD17} proposed cost-aware targeted viral marketing (CTVM) problem. They proposed Benefit Sampling Algorithm (BSA) based on RIS to generate random hyperedges. The difference is that the source node selected by BSA is proportional to the benefit of each node, while RIS selects randomly and uniformly. In addition, \cite{DBLP:journals/ton/NguyenTD17} uses Weighted-Max-Coverage algorithm and provides $(1-1 / \sqrt{e}-\epsilon)$ approximation ratio. \cite{DBLP:journals/ipl/KhullerMN99} proved that if the cost of the node is not uniform in the algorithm, the process returns $(1-1 / \sqrt{e})$-approximate coverage, otherwise it returns $(1-1 / e)$-approximate coverage. Guo \etal \cite{DBLP:conf/sigmod/Guo0WC20} proposed a new RR set Generation Scheme, which maps the selection of in-neighbors to subset sampling, and solves the IM problem through geometric distribution sampling. In geometric distribution sampling, each cycle can skip $i-1$ neighbor nodes to activate the $i-th$ neighbor node directly, instead of judging whether to activate each neighbor like the traditional method. They improve the time complexity to $O\left(k \cdot n \log n / \epsilon^{2}\right)$.
\vspace{-2ex}\subsection{Heuristic Algorithms}\label{ssec5.2}
Kempe \etal \cite{DBLP:conf/kdd/KempeKT03} also proposed a seed selection process guided by heuristics. Specifically, node degree and Pagerank score are utilized to guide the selection. Compared to approximate algorithms, such heuristic algorithms often exhibit excellent efficiency and scalability, but the quality of the results cannot be guaranteed. Since then many heuristic algorithms have been developed to address the IM problem. These algorithms can be generally classified into two categories, \textit{local} and \textit{global}, depending on how much knowledge the heuristics have to take into account. 

\vspace{-2ex}\subsubsection{Local Methods}
Since accurately computing the influence spread of a node (set) is $\#P$-hard, a number of heuristic approaches explore if the local topological structure of a network (\eg node degree) can be exploited to design score functions for each node to iteratively select nodes into the seeds set. Subsequently, nodes with high scores can be selected as seeds. As these score functions do not need to perform MC simulations or snapshot (\textit{resp}., RR sets) samples, the running time can be significantly saved compared to the approximate solutions.

Degree and other centrality-based heuristics are commonly used to estimate the influence of nodes in social networks \cite{DBLP:books/cu/WF1994}. Chen \etal proposed a heuristic method, \textsf{DegreeDiscount} \cite{DBLP:conf/kdd/ChenWY09}, based on node degree. If a node $u$ has been selected as a seed, the influence scores of its neighbors will be discounted. Specifically, given that the node $u$ is selected as a seed, during the next selection of a seed, the influence scores of $u's$ neighbors $v\in N_u=\{\{v\}|(u,v)\in E\}$ is subtracted by a specific value, which depends on the degree of $v$, the number of $v's$ neighbors, and a parameter $p$. In each round, it selects the node with the maximum influence score as the seed. Under the IC model, \textsf{DegreeDiscount} outperforms \textsf{CELF} \cite{DBLP:conf/kdd/LeskovecKGFVG07} and \textsf{NewGreedy} \cite{DBLP:conf/kdd/ChenWY09} in terms of running time, while preserving good result quality empirically.

Instead of only considering the degree of nodes, some works take into account more topological information. \cite{DBLP:conf/kdd/ChenWW10} restricts the information diffusion of a node $u$ to a local tree structure rooted at $u$. They introduce a model called \textsf{PMIA}, in which it first computes the \textit{maximum influence paths} (MIP) between every pair of nodes in the network via Dijkstra algorithm. It ignores MIPs with probability smaller than an influence threshold, effectively restricting influence to a local region. \textsf{PMIA} then unions the MIPs starting or ending at each node into arborescence structures that represent the local influence regions of each node. However, this model only considers influence propagated through these local arborescences. Surprisingly, experiments show that \textsf{PMIA} is close to the \textsf{Greedy} algorithm in terms of influence spread with excellent running time.

Kim \etal~ \cite{DBLP:conf/icde/KimKY13} proposed \textsf{IPA} under IC model, which estimates influence spread by considering an \textit{independent influence path} as an influence evaluation unit. Specifically, \textsf{IPA} treats an influence path from a seed node to a non-seed node as an influence evaluation unit and assumes that influence paths are independent of each other. Then, influence estimation only requires a series of simple arithmetic of influence paths. Since finding all influence paths between two nodes in tractable time is $\# P$-hard, \textsf{IPA} boosts the processing time for influence spread evaluation by controlling the number of influence paths using a pre-defined threshold. \eat{ Moreover, by considering an influence path as a fine-grained influence evaluation unit, most parts of \textsf{IPA} are parallelizable by organizing influence paths appropriatelyand adding a few lines of OpenMP expressions}.
\textsf{IPA} empirically provides comparable influence spread close to that of the Greedy solution obtained by Monte-Carlo simulations~\cite{DBLP:conf/kdd/KempeKT03}. Overall, it is superior to the influence spread computed by \textsf{PMIA}. However, \textsf{IPA} takes up more memory than \textsf{PMIA} as \textsf{IPA} stores all influence paths between any pair of nodes while \textsf{PMIA} stores only one path between them.

Since computing influence in directed acyclic graphs (DAGs) can be performed in linear time to the size of the graphs, Chen \etal proposed the first scalable influence maximization algorithm \textsf{LDAG} tailored for the LT model~\cite{DBLP:conf/icdm/ChenYZ10}. \textsf{LDAG} constructs local DAGs surrounding every node $u$ in the network and restricts the influence to $u$ to be within the DAG structure. This strategy is similar to \textsf{PMIA}. Specifically, there are two phases in \textsf{LDAG}. Given a threshold $\theta$, the first phase is to generate $LDAG(u,\theta)$ for every node $u$ using Dijkstra algorithm, which contains nodes that can influence $u$ with probabilities of at least $\theta$. Then based on the constructed local DAGs, the seed set can be obtained by greedily selecting nodes with maximum incremental influence. \textsf{LDAG} is reported to be orders of magnitude faster than Algorithm~\ref{alg:greedy}~\cite{DBLP:conf/kdd/KempeKT03}, and performs consistently among the best algorithms. On the other hand, other heuristic algorithms (\eg \textsf{DegreeDiscount}) are not specifically designed for the LT model and have unstable performances on different real-world networks~\cite{DBLP:journals/corr/LuX0HL17}.

There exist two limitations in \textsf{LDAG} \cite{DBLP:conf/icdm/ChenYZ10}. First, it relies heavily on finding a high-quality \textsf{LDAG}, which is in fact NP-hard. Hence, a greedy heuristic is employed to discover a good LDAG. As LDAG is already a heuristic, using a greedy LDAG in place of the optimal one may introduce additional loss in result quality of the seed set. Second, it assumes influence flows to a node via paths within only one LDAG and ignores influence via other paths. To overcome these limitations, \textsf{SIMPATH}~\cite{DBLP:conf/icdm/GoyalLL11} adopts the lazy update strategy introduced in~\textsf{CELF}~\cite{DBLP:conf/kdd/LeskovecKGFVG07}, while the spread can be computed by enumerating \textit{simple} paths (\ie a path in which nodes are repeated) starting from the seed nodes and restricting the enumeration to a small neighborhood by pruning paths with probability lower than a threshold. In \textsf{SIMPATH}, the spread of a node can be computed by summing up the weights (\ie probabilities) of the restricted simple paths originating from it.
The threshold affects the result quality since seed set quality is based on its spread of influence. The larger is the scope of its spread, the better is its quality. Empirically, \textsf{SIMPATH} outperforms \textsf{LDAG} in terms of time and space efficiencies. However, larger the size of seed set, more enumerated simple paths is required. In fact, \textsf{LDAG} is better than \textsf{SIMPATH} under certain cases as reported in~\cite{DBLP:journals/corr/LuX0HL17}.

\vspace{-2ex}\subsubsection{Global Methods}
Global heuristic methods take into account the topological characteristics of the entire network for computing the influence score of a node.

Kimura \etal \cite{DBLP:conf/pkdd/KimuraS06} defined a \textit{Shortest-Path Model} (\textsf{SPM}) to estimate the influence of nodes. Formally, let $d(u,v)$ be the distance from $u$ to $v$, then the distance for a node set $S$ to $v$ is defined as
$
d(S,v)=min_{u\in S}d(u,v).
$
Under \textsf{SPM}, each node $v$ has chance to become active only at step $t = d(\mathcal{X}_0, v)$. In other words, each node is activated exclusively through the shortest paths from the initial active node set. Furthermore, \textsf{SP1M} \cite{DBLP:conf/pkdd/KimuraS06} gives another chance for $v$ to be activated at step $t=d(\mathcal{X}_0, v)+1$. In this way, \textsf{SP1M} strictly limits the number of influence paths from a node set $S$ to any node $v$ by pruning all paths with lengths larger than $d(S,v)+1$. Thus the influence $\sigma(S)$ can be efficiently computed by Dijkstra algorithm.  However, a key limitation of this method is that it ignores the influence probability between nodes, which is important in real world applications.

\textsf{SPIN} \cite{DBLP:journals/tase/NarayanamN11} uses the \textit{Shapley value} in cooperative game theory~\cite{Shapley1953} to evaluate the marginal contribution of nodes in the influence propagation process. Since the time complexity for computing the Shapley value of each node is $O((n/e)^n)$~\cite{DBLP:journals/tase/NarayanamN11}, \textsf{SPIN} utilizes a sampling-based approach to approximate the Shapley value. \eat{Specifically, $t$ permutations of nodes in a network are randomly generated first and the influence contribution of each node is estimated afterwards. Finally, nodes are sorted in decreasing order of their average contribution values to construct the ranked list of nodes. To select $top$-$k$ nodes, \textsf{SPIN} selects the $top$-$1$ node in the ranked list as the first seed and adds nodes to the list of $top$-$(k-1)$ nodes that are not adjacent.} The advantage of \textsf{SPIN} is that its execution time is almost independent of the size of seed set, \ie $k$.
Jung \etal proposed \textit{Influence Ranking Influence Estimation} (\textsf{IRIE}) \cite{DBLP:conf/icdm/JungHC12}. \eat{When selecting the first seed, this method adopts a \textit{global influence ranking} (IR) method derived from a belief propagation approach, and then selects the highest-ranked node as the first seed. Afterwards, IR is combined with a simple influence estimation to evaluate the marginal influence of other nodes w.r.t this seed.} \eat{Next, \textsf{IRIE} uses the results to adjust the next round computation of influence ranking.} Specifically, \textsf{IRIE} derives the following system of linear equations for the influence spread ${\sigma(u|S,u\in V)}=(1-AP_S(u))\cdot (1+\alpha(\sum_{v\in N^{out}(u)}P_{u,v}\cdot r(v)))$,
where $r(u)$ is the influence of node $u$ in the graph, $S$ is the seed set, $\alpha \in (0,1)$ is a damping factor, and $AP_S(u)$ denotes the probability that node $u$ becomes activated after the diffusion process, which can be estimated by \textsf{PMIA} and Monte Carlo simulations. Experiments show that \textsf{IRIE} shows high accuracy and is extremely efficient. Moreover, compared to \textsf{PMIA}, which needs to store data structures related to the local influence region of every node, \textsf{IRIE} uses global iterative computations without storing additional data structures. Hence, the memory overhead is smaller.

Liu \etal \cite{DBLP:conf/cikm/LiuXCXTY14} proposed a method called \textsf{GroupPR}, which adopts a linear and tractable approach to describe the influence propagation \cite{DBLP:conf/ijcai/XiangLCXZY13} and uses a quantitative metric based on PageRank (\textit{Group-PageRank}) to quickly estimate the upper bound of the social influence for any node set $S$. They presented a method that can compute \textit{Group-PageRank} for each node in $O(m)$ time. 
%

\textsf{IMRANK} \cite{DBLP:conf/sigir/ChengSHCC14} intends to get ranks of nodes consistent with their marginal influence spread so that it selects the $top$-$k$ nodes in the rank as seeds. Specifically, it starts from an initial rank of nodes obtained from an efficient heuristic algorithm and reorders nodes based on their marginal influence spread with respect to the current ranking. After a finite number of iterations, \textsf{IMRANK} converges to a consistent ranking. \eat{However, in each iteration, estimating the marginal influence of all nodes w.r.t the current ranking is expensive. Hence, \textsf{IMRANK} uses a \textit{last-to-first allocating} (LFA) strategy to improve the efficiency. The key idea of LFA is based on two rules: (1) each node can only be activated by nodes ranked higher than it in the given ranking; (2) when a node could be activated by multiple nodes, the higher-ranked node has a higher priority to activate it.} Experiments show that compared to the greedy algorithm~\cite{DBLP:conf/kdd/KempeKT03}, \textsf{IMRANK}, with a ``good'' initial ranking shows indistinguishable performance. In terms of efficiency, \textsf{IMRANK} outperforms other heuristic algorithms such as \textsf{PMIA} \cite{DBLP:conf/kdd/ChenWW10} and \textsf{IRIE}\cite{DBLP:conf/icdm/JungHC12}.

Galhotra \etal~\cite{DBLP:conf/sigmod/GalhotraAR16} proposed a heuristic algorithm named \textsf{EasyIM}. The key idea behind this technique is that a simple function $F^l(u)$, the number of paths from a node $u$ to all other nodes $v \in V\setminus\{u\}$, can be used to assign a score to $u$, where $l$ denotes the maximum length of paths. 

\begin{table*}[!]
	\centering\vspace{-2ex}
	\caption{Comparison of Classical IM techniques.}
	\label{tab:IM_works}\vspace{-2ex}
	\resizebox{\textwidth}{!}{
		\begin{threeparttable}
			\begin{tabular}{|c|lllcccc|}
				\toprule  
				Category& Subcategory& Method & Infor. Diffu. models& Expected Complexity & Worst-case Complexity &Space Complexity& Approximation\\
				\midrule  
				\rule{0pt}{15pt}
				\multirow{14}{0.8in}{\shortstack{Approximate}} & \multirow{5}{0.8in}{\tabincell{c}{Simulation\\-based}}
				\rule{0pt}{15pt}
				& \textsf{Greedy} \cite{DBLP:conf/kdd/KempeKT03} &IC,LT,CT&$O(krn\sigma)$&$O(krnm)$ & $O(n\sigma)$&$1-1/e-\varepsilon(r)$   \\
				\rule{0pt}{15pt}
				&&\textsf{CELF} \cite{DBLP:conf/kdd/LeskovecKGFVG07}&IC,LT,CT&$O(krn'\sigma)$&$O(krnm)$& $O(n\sigma)$&$1-1/e-\varepsilon(r)$ \\
				
				\rule{0pt}{15pt}
				&&\textsf{CGA} \cite{DBLP:conf/kdd/WangCSX10}&IC& &\tabincell{c}{$O(m)+O(rm_p)\cdot$ \\$O(n(Z-M)+k(M+n_p))$}& $O(n+k+M)$&$1-e^{\frac{1}{1+\delta \rho}}$ \\
				\rule{0pt}{15pt}
				&&\textsf{CELF++} \cite{DBLP:conf/www/GoyalLL11}&IC,LT,CT&$O(krn'\sigma)$&$O(krnm)$& $O(n\sigma)$&$1-1/e-\varepsilon(r)$ \\
				\rule{0pt}{15pt}
				&&\textsf{UBLF} \cite{DBLP:journals/tkde/ZhouZZG15}&IC,LT,CT&$O(krn'\sigma)$& $O(krnm)$&$O(n^2\sigma)$&$1-1/e-\varepsilon(r)$ \\
				
				\cmidrule{2-8}
				
				& \multirow{9}{0.8in}{\tabincell{c}{Sampling\\-based}}
				\rule{0pt}{15pt}
				&\textsf{NewGreedy} \cite{DBLP:conf/kdd/ChenWY09}&IC& &$O(krnm)$& $O(rm')$&$1-1/e-\varepsilon(r)$ \\
				\rule{0pt}{15pt}
				&&\textsf{StaticGreedyDU} \cite{DBLP:conf/cikm/ChengSHZC13}&IC& &$O\left(\frac{kmn^2log\dbinom{n}{k}}{\varepsilon^2}\right)$ & $O(rm')$&$1-1/e-\varepsilon(r)$  \\
				\rule{0pt}{15pt}
				&&\textsf{SKIM} \cite{DBLP:conf/cikm/CohenDPW14}&IC& &$O\left(\frac{kmn^2log\dbinom{n}{k}}{\varepsilon^2}\right)$& $O(|R'|m')$&$1-1/e-\varepsilon$ \\
				\rule{0pt}{15pt}
				&&\textsf{PMC} \cite{DBLP:conf/aaai/OhsakaAYK14}&IC& &$O\left(\frac{kmn^2log\dbinom{n}{k}}{\varepsilon^2}\right)$& $O(|R_{DAG}|(m'+|V'|^2))$&$1-1/e-\varepsilon$ \\
				\rule{0pt}{15pt}
				&&\textsf{RIS} \cite{DBLP:conf/soda/BorgsBCL14}&IC,LT,CT& $O(\frac{kl^2(m+n)log^2n}{\varepsilon^3})$& & $O(|R_\theta||R|)$&$1-1/e-\varepsilon$ \\
				\rule{0pt}{15pt}
				&&\textsf{TIM}/\textsf{TIM}$^+$ \cite{DBLP:conf/sigmod/TangXS14}&IC,LT,CT& $O(\frac{(k+l)(m+n)logn}{\varepsilon^2})$& & $O(|R_\theta||R|)$&$1-1/e-\varepsilon$ \\
				\rule{0pt}{15pt}
				&&\textsf{IMM} \cite{DBLP:conf/sigmod/TangSX15}&IC,LT,CT& $O(\frac{(k+l+\gamma)(m+n)logn}{\varepsilon^2})$& & $O(|R_\theta||R|)$&$1-1/e-\varepsilon$ \\
				\rule{0pt}{15pt}
				&&\textsf{SSA}/\textsf{DSSA} \cite{DBLP:conf/sigmod/NguyenTD16}&IC,LT,CT& & & $O(|R_\theta||R|)$&$1-1/e-\varepsilon$ \\
				\rule{0pt}{15pt}
				&&\textsf{BKRIS} \cite{DBLP:journals/tkde/WangZZLC17}&IC,LT& \tabincell{c}{$O\left(\frac{|R|k(m+n)}{\varepsilon^2}\right)$$\cdot$ \\ $O\left(logn+log\dbinom{n}{k}\right)$} &$O\left(\frac{mn(logn+log\dbinom{n}{k})}{\varepsilon^2}\right)$ & $O(|R_\theta||R|)$&$1-1/e-\varepsilon-\varepsilon'$ \\
				\midrule
				\rule{0pt}{15pt}
				\multirow{13}{0.8in}{Heuristic} & \multirow{5}{0.8in}{\tabincell{c}{Local}}
				\rule{0pt}{15pt}
				&\textsf{DegreeDiscount} \cite{DBLP:conf/kdd/ChenWY09}&IC,LT,CT& &$O(klogn+m)$& $O(n)$& \\
				\rule{0pt}{15pt}
				&&\textsf{LDAG} \cite{DBLP:conf/icdm/ChenYZ10}&LT& &\tabincell{c}{$O(nt_\theta)+$\\ $O(kn_\theta m_\theta(m_\theta+logn))$}& $O(n\overline{m}_\theta)$&  \\
				\rule{0pt}{15pt}
				&&\textsf{PMIA} \cite{DBLP:conf/kdd/ChenWW10}&IC& &\tabincell{c}{$O(nt_{i\theta})$+\\$O(kn_{0\theta} n_{i\theta}(n_{i\theta}+logn))$}& $O(n(n_{i\theta}+n_{0\theta}))$& \\
				\rule{0pt}{15pt}
				&&\textsf{SIMPATH} \cite{DBLP:conf/icdm/GoyalLL11}&LT& &$O(lknP_\theta)$& $O(n\overline{m})$& \\
				\rule{0pt}{15pt}
				&&\textsf{IPA} \cite{DBLP:conf/icde/KimKY13}&IC& &\tabincell{c}{$O(\frac{nO_vn_{vu}}{c})+$\\ $O(k^2(\frac{O_vn_{vu}}{c}+(c-1)))$}& $O(3k\overline{m})$& \\
				
				\cmidrule{2-8}
				\rule{0pt}{15pt}
				& \multirow{8}{0.8in}{\tabincell{c}{Global}}
				\rule{0pt}{15pt}
				&\textsf{SP1M} \cite{DBLP:conf/pkdd/KimuraS06}&IC& &$O(knm)$& $O(n+m+k)$& \\
				\rule{0pt}{15pt}
				&&\textsf{IRIE} \cite{DBLP:conf/icdm/JungHC12}&IC& &$O(k(n_{0\theta}k+m)))$& $O(n)$& \\
				\rule{0pt}{15pt}
				&&\textsf{BPPB} \cite{DBLP:journals/kais/SaitoKOM12}&SIS& $O(MTm)$& & $O(T(n+m)+(T-1)m)$& \\
				\rule{0pt}{15pt}
				&&\textsf{SPIN} \cite{DBLP:journals/tase/NarayanamN11}&LT& &\tabincell{c}{$O(tr(m+n))+$\\$O(nlogn+kn)$}& $O(n)$& \\
				\rule{0pt}{15pt}
				&&\textsf{SVIM} \cite{DBLP:conf/wsdm/Li0WZ13}&Voter&
				\tabincell{c}{$\begin{cases}
					O(t\cdot|E|) & \textsf{SVIM-S} \\ O(|E|+min(bn_z^3+n_x^3,t_Cm_B))  & \textsf{SVIM-L}
					\end{cases}$} & & \tabincell{c}{$\begin{cases}
					O(n^2) & \textsf{SVIM-S} \\ O(m_Bn^2)  & \textsf{SVIM-L}
					\end{cases}$}& \\
				\rule{0pt}{15pt}
				&&\textsf{GroupPR} \cite{DBLP:conf/cikm/LiuXCXTY14}&IC&
				&\tabincell{c}{$\begin{cases}
					O(knm)& \text{Linear}\\
					O(m+k^2n)& \text{Bound}
					\end{cases}
					$} & $O(n)$& \\
				\rule{0pt}{15pt}
				&&\textsf{IMRANK} \cite{DBLP:conf/sigir/ChengSHCC14}&IC& &$O(nTd_{max}logd_{max})$& $O(n)$& \\
				\rule{0pt}{15pt}
				&&\textsf{EasyIM} \cite{DBLP:conf/sigmod/GalhotraAR16}&IC,LT& &$O(Dk(m+n))$& $O(n)$&  \\
				\bottomrule 
			\end{tabular}
			
			\begin{tablenotes}
				\footnotesize
				\item[1] $r$ is the number of rounds for MC simulations, $O(\sigma)$ denotes the expected complexity of a simulation.$n’$ is the number of simulation at each round. $n_p,m_p$ are the number of nodes and edges in the largest community; $Z,M$ are the number of communities before and after the combination; $\varepsilon(r)$ denotes the error using $r$ MC simulations/snapshots to evaluate the influence of nodes, and the value of $r$ is given in advance.
				\item[2] $m’$ denotes the average number of edges in every sampling graph; $|R'|$ is the number of the sampling graphs; $|R_{DAG}|$ denotes the number of DAGs generated in PMC;  $|V'|$ denotes the average number of nodes in decomposed sampling graph;  $|R_\theta|$ denotes the number of generated RR sets; $|R|$ is the expected size of an RR set; $l$ is a user-specified factor that controls the trade-off between result quality and efficiency, and usually set to 1; $\gamma$ is a  small constant and is set to 2.5 in IMM.
				\item[3] $t_\theta$ is the time complexity for constructing the LDAG for one node, $n_\theta$ and $m_\theta$ are the maximum number of nodes and edges in one LDAG; $t_{i\theta}$ is the time complexity for constructing the MIIA for one node, $n_{i\theta}$ and $n_{0\theta}$ are the maximum size of one single MIIA and MIOA; $P_\theta$ is the maximum number of simple paths starting from one node, $l$ is the look head value;  $O_v$ is the maximum number of nodes having influence paths from one node; $n_{vu}$ is the maximum number of influence paths between any two nodes; $c$ is the number of parallelized processes; $\overline{m}_\theta$ denotes the average size in terms of the number of edges in $LDAG(v,\theta)$; $\overline{m}$ denotes average number of simple paths starting from one node.
				\item[4] $n_{0\theta}$ is the maximum size of one single MIOA; $M$ is the times of BP process, $T$ is the number of time step; $t$ is the size of the sampled set from $G$; $b$ is the number of balanced sink components in $G$, $n_z$ is the number of nodes in the largest balanced sink component,
				$n_x=|x|$, where $x = V \setminus z$, $t_C$ is the largest convergence time,
				$m_B$ is the number of edges of the induced subgraph $G_B$ consisting of all nodes in the balanced sink components and $x$;
				$T$ is the number of iterations before convergence; $d_{max}$ is the largest number of paths end in a node with length no more than $l$ (usually $l=1$); $D$ is the edge depth of $G$.
			\end{tablenotes}
		\end{threeparttable}
	}\vspace{-4ex}
\end{table*}

There also exist some works on non-progressive diffusion models. For instance, Kazumi \etal \cite{DBLP:journals/kais/SaitoKOM12} proposed an effective method using \textit{bond percolation} (\textsf{BP})~\cite{DBLP:journals/datamine/KimuraSNM10} with \textit{pruning} and \textit{burnout} (\textsf{BPPB}) to the IM problem under the SIS model. \eat{Considering the impact of time, \textsf{BPPB} first generates a layer graph $G^T=(V^T,E^T)$ from $G$ in the following way. First, for each node $v\in V$ and each time-step $t\in \{0, 1,\ldots, T \}$, $v_t$ is a copy of $v$ at time-step $t$, and $E_t = \{(u_{t-1}, v_t)| (u, v) \in E\}, E^T = E_1 \cup\ldots\cup E_T$. \textsf{BP} process in graph $G^T$ is a random process where each link $(u_{t-1},v) \in E^T$ is independently declared ``occupied'' with probability $p_{u_{t-1},v}$. Based on it, \textsf{BP} simultaneously estimates the influence spread for all $v\in V\verb|\| S$.} 
In order to improve the efficiency, \textsf{BPPB} combines a \textit{pruning}~\cite{DBLP:conf/ijcai/KimuraSM09} and \textit{burnout}~\cite{DBLP:conf/dis/SaitoKM09} with \textsf{BP}. \eat{The key idea of the pruning method is that once $H(S \cup \{u\}, t_0) = H(S \cup \{v\}, t_0)$ at some time-step $t_0$ on the course of the BP process for a pair of information source nodes, $u$ and $v$, the $H(S\cup \{u\}, t)$ equals to $H(S \cup \{v\}, t)$  for all $t \ge t_0$, where $H(S,t)$ denotes the set of activated nodes at time $t$. Moreover, with burnout technique, \textsf{BPPB} substantially reduce the size of the active node set from $H(S \cup \{v\}, t)$ to $H(\{v\}, t)$ for each $v \in V \verb|\|S$ and $t \in \{1,\ldots, T \}$ compared to \textsf{BP} with pruning method.} Besides,
Eyal \etal \cite{DBLP:journals/ipl/Even-DarS11} conducted a comprehensive study for the influence maximization problem under the Voter model. They justified that given $S = \{u : u\in\mathcal{X}_0\}$ (according to Eq.~\ref{eq3}), the probability for $v\in\mathcal{X}_t$, \ie $p[v\in\mathcal{X}_t]$, can be interpreted as the probability for a random walk of length $t$ starting at $v$ ending in $S$. Based on this finding, they studied the case of short term and long term influence. Li \etal \cite{DBLP:conf/wsdm/Li0WZ13} proposed methods \textsf{SVIM-S} and \textsf{SVIM-L} to solve the IM problem in a signed graph under the Voter model.

\vspace{-2ex}\subsection{Summary}
Table~\ref{tab:IM_works} summarizes key features of the classical IM techniques discussed in this section. Specifically, it compares the supported diffusion models, time complexity (including both expected and worst-case ones), space complexity, as well as the approximation ratio over the results quality (if any) for these techniques. In summary, the approximate solutions to the IM problem can theoretically provide guarantee over the result quality but suffer from huge computational cost. In contrast, the heuristic methods sacrifice the guarantee over results quality for efficiency. Within approximate solutions, latest works on RR set-based samplings can achieve comparable efficiency with heuristic solutions. On the other hand, state-of-the-art heuristic solutions can empirically achieve comparable result quality with respect to approximate ones, although they fail to provide theoretical guarantee.

In summary, some of these techniques can discover seeds with a theoretical guarantee, some are effective in practical result quality, some are efficient in terms of running time or space overhead. Unfortunately, there is no method that is dominant in all the fundamental performance criteria (\ie result quality, running time, and space overhead). Therefore, an elegant method that can address classical IM by performing superiorly for all these three criteria remains an open problem.

\begin{figure*}[t]
	\begin{tikzpicture}[xscale=0.5, yscale=0.31, only marks, y=.5cm,
	topic/.style={circle, draw=blue, fill=blue,minimum size=0.4cm},
	social/.style={rectangle, draw=orange, fill=orange,minimum size=0.4cm},
	dynamic/.style={diamond, draw=magenta, fill=magenta,minimum size=0.4cm},
	time/.style={star,draw=olive, fill=olive,minimum size=0.4cm},
	competitive/.style={regular polygon,regular polygon sides=5, draw=red, fill=red,minimum size=0.45cm},
	location/.style={isosceles triangle, draw=teal, fill=teal,minimum size=0.4cm},
	budget/.style={semicircle, draw=black, fill=black, minimum size=0.3cm},
	adaptive/.style={star,star points=3, draw=red!25!blue, fill=red!25!blue,minimum size=0.4cm}]
	\tikzstyle{every node}=[font=\footnotesize,xscale=0.5,yscale=0.41]
	
	\draw[->,line width=.05cm,xshift=0cm] (0,0) -- coordinate (x axis mid) (22,0);
	\draw[dash pattern= on 0.25cm off 0.25cm,line width=.05cm,xshift=0cm] (0,0) -- coordinate (y axis mid) (0,25);
	\foreach \x/\xtext in {0/2006,1.5/2007,3/2008,4.5/2009,6/2010,7.5/2011,9/2012,10.5/2013,12/2014,13.5/2015,15/2016,16.5/2017,18/2018,19.5/2019,21/2020}
	\draw [xshift=0cm](\x cm,1.6pt) -- (\x cm,-3pt)
	node[anchor=north,font=\Large] {$\xtext$};
	
	\node (1-09) [topic,label={[black!100,font=\Large]above:\cite{DBLP:conf/kdd/TangSWY09}}] at (4.5,1) {};
	\node (1-13) [topic,label={[black!100,font=\Large]above:\cite{DBLP:conf/cikm/GuoZZCG13}}] at (10.5,1) {};
	\node (1-14) [topic,label={[black!100,font=\Large]above:\cite{DBLP:conf/edbt/AslayBBB14}}] at (12,1) {};
	\node (1-15) [topic,label={[black!100,font=\Large]above:\cite{DBLP:journals/pvldb/ChenFLFTT15,DBLP:journals/pvldb/LiZT15}}] at (13.5,1) {};
	\node (1-16) [topic,label={[black!100,font=\Large]above:\cite{DBLP:conf/infocom/NguyenDT16}}] at (15,1) {};
	\node (1-17) [topic,label={[black!100,font=\Large]above:\cite{DBLP:conf/sigmod/LiFZT17}}] at (16.5,1) {};
	\node (1-18) [topic,label={[black!100,font=\Large]above:\cite{DBLP:conf/icde/FanQLMZLTD18}}] at (18,1) {};
	\node (1-19) [topic,label={[black!100,font=\Large]above:\cite{DBLP:journals/isci/ChenQYLS20}}] at (21,1) {};
	
	\node (2-14) [location,label={[black!100,font=\Large]above:\cite{DBLP:conf/sigmod/LiCFTL14}}] at (12,4) {};
	\node (2-15) [location,label={[black!100,font=\Large]above:\cite{DBLP:conf/cikm/ZhouCLXZL15,DBLP:conf/kdd/ZhuPCZZ15}}] at (13.5,4) {};
	\node (2-16) [location,label={[black!100,font=\Large]above:\cite{DBLP:conf/cikm/SongHL16,DBLP:conf/icde/WangZZL16}}] at (15,4) {};
	\node (2-17) [location,label={[black!100,font=\Large]above:\cite{DBLP:journals/tkde/GuoZCWT17,DBLP:journals/tkde/WangZZL17}}] at (16.5,4) {};
	\node (2-18) [location,label={[black!100,font=\Large]above:\cite{DBLP:conf/dasfaa/0002ZZLQ18}}] at (18,4) {};
	\node (2-19) [location,label={[black!100,font=\Large]above:\cite{DBLP:journals/kais/SaleemKCP19}}] at (19.5,4) {};
	
	\node (3-12) [dynamic,label={[black!100,font=\Large]above:\cite{DBLP:conf/sdm/AggarwalLY12}}] at (9,7) {};
	\node (3-13) [dynamic,label={[black!100,font=\Large]above:\cite{DBLP:conf/icdm/ZhuangSTZS13}}] at (10.5,7) {};
	\node (3-15) [dynamic,label={[black!100,font=\Large]above:\cite{DBLP:conf/sdm/ChenSHX15}}] at (13.5,7) {};
	\node (3-16) [dynamic,label={[black!100,font=\Large]above:\cite{DBLP:journals/pvldb/OhsakaAYK16}}] at (15,7) {};
	\node (3-17) [dynamic,label={[black!100,font=\Large]above:\cite{DBLP:journals/pvldb/WangFLT17}}] at (16.5,7) {};
	\node (3-18) [dynamic,label={[black!100,font=\Large]above:\cite{DBLP:journals/tois/WangLFT18}}] at (17.8,7) {};
	\node (3-19) [dynamic,label={[black!100,font=\Large]above:\cite{DBLP:journals/geoinformatica/WangLYH19,DBLP:conf/icde/ZhaoSWLZ19,DBLP:conf/cikm/0001WJPC19}}] at (19.5,7) {};
	
	\node (4-09) [time,label={[black!100,font=\Large]above:\cite{DBLP:conf/acml/SaitoKOM09}}] at (4.5,10) {};
	\node (4-10) [time,label={[black!100,font=\Large]above:\cite{DBLP:conf/kdd/ChenWW10}}] at (6,10) {};
	\node (4-11) [time,label={[black!100,font=\Large]above:\cite{DBLP:conf/icml/Gomez-RodriguezBS11}}] at (7.5,10) {};
	\node (4-12) [time,label={[black!100,font=\Large]above:\cite{DBLP:conf/aaai/ChenLZ12,DBLP:conf/icml/Gomez-RodriguezS12,DBLP:conf/icdm/LiuCXZ12}}] at (9,10) {};
	\node (4-13) [time,label={[black!100,font=\Large]above:\cite{DBLP:conf/nips/DuSGZ13}}] at (10.5,10) {};
	\node (4-16) [time,label={[black!100,font=\Large]above:\cite{DBLP:conf/pkdd/OhsakaYKK16}}] at (15,10) {};
	\node (4-19) [time,label={[black!100,font=\Large]above:\cite{DBLP:conf/icde/HuangBCZ19}}] at (19.5,10) {};
	
	\node (5-13) [social,label={[black!100,font=\Large]above:\cite{DBLP:conf/edbt/LiBS13}}] at (10.5,13) {};
	\node (5-15) [social,label={[black!100,font=\Large]above:\cite{DBLP:journals/vldb/0005BSC15}}] at (13.5,13) {};
	\node (5-18) [social,label={[black!100,font=\Large]above:\cite{DBLP:conf/wcsp/LiGFTQZ18}}] at (18,13) {};
	
	\node (6-07) [competitive,label={[black!100,font=\Large]above:\cite{DBLP:conf/wine/BharathiKS07,DBLP:conf/ACMicec/CarnesNWZ07}}] at (1.5,16) {};
	\node (6-10) [competitive,label={[black!100,font=\Large]above:\cite{DBLP:conf/wine/BorodinFO10}}] at (6,16) {};
	\node (6-11) [competitive,label={[black!100,font=\Large]above:\cite{DBLP:conf/www/BudakAA11,DBLP:conf/gamesec/ClarkP11}}] at (7.5,16) {};
	\node (6-12) [competitive,label={[black!100,font=\Large]above:\cite{DBLP:conf/sdm/HeSCJ12}}] at (9,16) {};
	\node (6-15) [competitive,label={[black!100,font=\Large,above left=.02cm]:\cite{DBLP:conf/sigmod/LiBCGM15,DBLP:conf/kdd/LinLC15,DBLP:journals/pvldb/LuCL15}}] at (13.5,16) {};
	\node (6-16) [competitive,label={[black!100,font=\Large]above:\cite{DBLP:conf/cikm/OuCC16,DBLP:conf/infocom/ZhuLZ16}}] at (15,16) {};
	\node (6-19) [competitive,label={[black!100,font=\Large]above:\cite{DBLP:journals/www/YanZLY19}}] at (19.5,16) {};
	\node (6-20) [competitive,label={[black!100,font=\Large]above:\cite{DBLP:journals/corr/abs-2012-03354}}] at (21,16) {};
	
	\node (7-11) [adaptive,label={[black!100,font=\Large]above:\cite{DBLP:journals/jair/GolovinK11}}] at (7.5,22) {};
	\node (7-13) [adaptive,label={[black!100,font=\Large]above:\cite{DBLP:conf/icml/ChenWY13}}] at (10.5,22) {};
	\node (7-15) [adaptive,label={[black!100,font=\Large]above:\cite{DBLP:conf/kdd/LeiMMCS15,DBLP:journals/corr/VaswaniL15}}] at (13.5,22) {};
	\node (7-16) [adaptive,label={[black!100,font=\Large]above:\cite{DBLP:journals/corr/VaswaniL16}}] at (15,22) {};
	\node (7-17) [adaptive,label={[black!100,font=\Large,above=.4cm]:\cite{DBLP:conf/icdm/LagreeCCM17,DBLP:conf/nips/WenKVV17,DBLP:conf/ijcai/YuanT17}}] at (16.5,22) {};
	\node (7-18) [adaptive,label={[black!100,font=\Large]above:\cite{DBLP:journals/pvldb/HanHXTST18,DBLP:conf/asunam/SalhaTV18,DBLP:conf/kdd/SunHYC18}}] at (18,22) {};
	\node (7-19) [adaptive,label={[black!100,font=\Large,above=.4cm]:\cite{DBLP:journals/tkdd/LagreeCCM19,DBLP:conf/sigmod/TangHXLTSL19,DBLP:conf/nips/PengC19}}] at (19.5,22) {};
	
	\node (8-13) [budget,label={[black!100,font=\Large]above:\cite{DBLP:journals/jsac/NguyenZ13}}] at (10.5,19) {};
	\node (8-14) [budget,label={[black!100,font=\Large]above:\cite{DBLP:conf/pakdd/HanZHS14}}] at (12,19) {};
	\node (8-20) [budget,label={[black!100,font=\Large]above:\cite{DBLP:journals/pvldb/BianGWY20, DBLP:conf/icde/Huang0XSL20}}] at (21,19) {};

	\node[below=1cm,location,label={[black!100,font=\Large]right:Location-aware IM}] at (1,0) {};
	\node[below=1cm,budget,label={[black!100,font=\Large]right:Budgeted IM}] at (6,0) {};
	\node[below=1cm,adaptive,label={[black!100,font=\Large]right:Adaptive IM}] at (11,0) {};
	\node[below=1cm,dynamic,label={[black!100,font=\Large]right:Dynamic-aware IM}] at (16,0) {};
	
	\node[below=1.7cm,time,label={[black!100,font=\Large]right:Time-aware IM}] at (1,0) {};
	\node[below=1.7cm,competitive,label={[black!100,font=\Large]right:Competitive IM}] at (6,0) {};
	\node[below=1.7cm,topic,label={[black!100,font=\Large]right:Topic-aware IM}] at (11,0) {};
	\node[below=1.7cm,social,label={[black!100,font=\Large]right:Social-psychology-aware IM}] at (16,0) {};
	\end{tikzpicture}
		\vspace{-2ex}\caption{Representative work and  milestones for extended IM solutions.}
		\label{fig:Extended_IMwork}\vspace{-4ex}
\end{figure*}
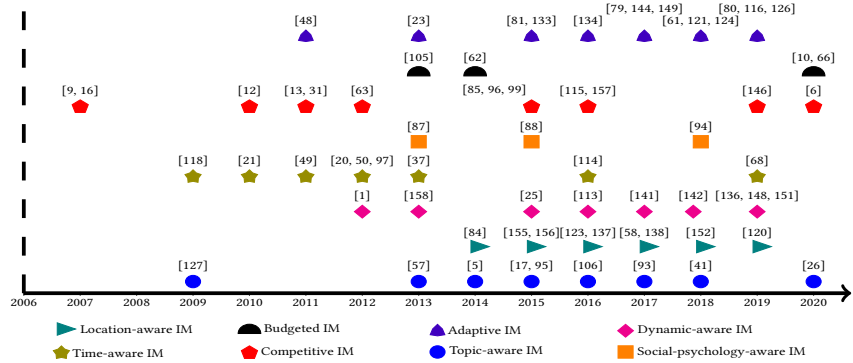

\vspace{-1ex}\section{Extended Influence Maximization}\label{sec6}
The classical IM works provide a fundamental mathematical model where influence propagation is viewed as a random process over a specific network without considering other factors that may be present in practical scenarios. In reality, in many practical social networks, the information diffusion process not only depends on the topology but also affected by other factors such as geo-location or topics of interest of users. Consequently, there have been many efforts that extend the classical IM model by considering some of these practical factors. We refer to these works collectively as \textit{extended} IM. Depending on the factors that are additionally taken into account with respect to the classical IM model, we classify these extended IM techniques into seven groups (recall our taxonomy in Section~\ref{sec2}).  Fig.~\ref{fig:Extended_IMwork} depicts the developmental road-map of these groups.

\vspace{-2ex}\subsection{Topic-aware IM}

In real-world social networks, a user may be interested in various topics \eat{(Fig.~\ref{fig:topic-aware})}. The topic a user $u$ favors will greatly affect the probability that $u$ is activated during the information diffusion process. Intuitively, users are more likely to be affected by other users with similar interests and information that is of interest to them. Recent research \cite{DBLP:conf/kdd/TangSWY09,DBLP:conf/cikm/GuoZZCG13,DBLP:journals/pvldb/ChenFLFTT15} shows that the classical IM problem cannot be trivially adapted to take into account the topic factor. Therefore, the problem of maximizing influence with respect to a particular topic, referred to as \textit{Topic-Aware Influence Maximization} (TIM), is introduced and studied extensively. \eat{This model focuses on maximizing the influence of users associated with the topic of a query under a particular diffusion model. Formally, let $\tau$ denote a query topic. Given a query $Q = (\tau, k)$, $\sigma_G (S | \tau)$ denotes the influence spread of the seed set $S$ in network $G$ when the query topic is $\tau$. The TIM problem aims to find $S$ with $k$ users by maximizing the targeted influence spread on other users in $G$ under the topic $\tau$.}
Barbieri \etal~\cite{DBLP:journals/kais/BarbieriBM13} extended the IC model to the \textit{Topic-aware Independent Cascade} (TIC) model. \eat{The strength of the relationship between two nodes according to this model is computed from the topic preferences learned from the historical activities in a social network. They also devised methods to learn model parameters from the log of past propagations.}
The main idea of the TIC model is to compute a probability $p$ for each edge $(u, v)\in E$ and each topic $z\in [1, Z]$ in a network. Based on it we can compute a probability $p_{u,v}^{z}$ representing the strength of the influence exerted by user $v$ on user $u$ w.r.t topic $z$. \eat{For each item $i$ that spreads across the network, we have a topic distribution, which is for each topic $z\in [1, Z]$, we are given $\gamma_{i}^{z}$, with $\sum_{z=1}^{Z}\gamma_{i}^{z}=1$. In the TIC model, propagation occurs similar to the IC model, \ie when a node $u$ becomes active under the influence of item $i$, it has one chance of influencing each inactive neighbor $v$, independent of the history thus far, with a probability which is the weighted average of the link probability w.r.t. the topic distribution of $i$:
$
p_{u,v}^{i}=\sum_{z=1}^{Z}\gamma_{i}^{z}p_{u,v}^{z}.
$
}
\vspace{-1ex}\subsubsection{IM for topic-dependent diffusion}
Inspired by~\cite{DBLP:journals/kais/BarbieriBM13}, recent research efforts~\cite{DBLP:conf/edbt/AslayBBB14,DBLP:journals/corr/ChenLY14,DBLP:conf/sigmod/LiFZT17} focus on associating a user's edge with a topic distribution. We refer to this type of method as \textit{IM for topic-dependent diffusion}, which associates each edge with the probability under each topic, evaluates the probability of propagation under each query, and tends to maximize influence under the new diffusion model. The main challenge here is the efficiency, as the distributions over topics will produce different probability maps for a large number of potential queries, making the computation of influence spread more expensive than traditional IM.
Aslay \etal~\cite{DBLP:conf/edbt/AslayBBB14} proposed \textsf{INFLEX} algorithm that solves the TIM problem under TIC by establishing a tree-based index for similarity search with Bregman divergences, which effectively improves query performance. 
Chen \etal~\cite{DBLP:journals/corr/ChenLY14} introduced a preprocessing strategy called \textsf{MIS} to pre-compute some seed sets for each topic and then aggregate them for online seed selection. Both \textsf{INFLEX} and \textsf{MIS} use samples to answer TIM queries directly. They are heuristic algorithms with no theoretical guarantees.

Besides TIC model, Chen \etal~\cite{DBLP:journals/pvldb/ChenFLFTT15} exploited the maximum influence arborescence (MIA) model to approximate the influence propagation in TIM problem. By estimating the upper limit of each user's topic-aware ability, this method preferentially calculates the marginal influence for users with a larger upper limit and prunes users with less influence to achieve influence spread with an approximation ratio of $(1-1/e)$. \eat{They further proposed three methods for boundary estimation, which are based on pre-calculation, local graph, and neighborhood-based, respectively.
They also designed a \textit{topic-sample} approach that pre-calculates seed sets of a set of sampled topic distributions and then uses the materialized topic distributions and previously calculated seed sets to estimate upper and lower bounds.}
Ju \etal~\cite{DBLP:conf/icde/FanQLMZLTD18} implemented an online topic-aware social influence analysis system called \textsf{OCTOPUS}. It mainly consists of three functions as follows: 1) a user-friendly influence analysis interface that allows users to use keywords to perform influence analysis, 2) three keyword-based topic-aware influence analysis tools, and 3) an online keyword query function, which is supported by the technology proposed in \cite{DBLP:journals/pvldb/ChenFLFTT15,DBLP:conf/sigmod/LiFZT17}.

\vspace{-2ex}\subsubsection{IM for topic-relevant targets}
The aforementioned strategies extend the classical IM problem to topic-aware IM by making the influence propagation model topic-aware. In particular, the propagation model needs to consider the probability of influence based on different topics. Consequently, this can adversely impact the efficiency of the solutions. A new set of methods emerged to address this challenge, which we refer to as \textit{IM for topic-relevant targets}~\cite{DBLP:conf/cikm/GuoZZCG13,DBLP:journals/pvldb/LiZT15,DBLP:conf/infocom/NguyenDT16}. The main idea is to consider a user as topic-aware and aim to maximize the influence spread on a subset of users related to the query topic. These studies mainly focus on distinguishing users and calculating the expected influence based on users who are interested in a query topic.

The \textit{Keyword-Based Targeted influence Maximization} (KB-TIM) problem in \cite{DBLP:journals/pvldb/LiZT15} creates a universal topic space for user interests and assigns each node $v$  a weighted term vector that describes a user's preferences on different topics. The topic for each query is also modeled as a weighted term vector. The relevance of a query topic and a user is then calculated as the similarity between the two vectors. 
\cite{DBLP:journals/isci/ChenQYLS20} proposed a Generalized Reverse Influence Set based framework for Semantics-aware Influence Maximization (GRIS-SIM). GRIS-SIM can determine how to generate RR sets by setting external sampling strategies, so that RR sets have different generation probabilities and weights.

Nguyen \etal~\cite{DBLP:conf/infocom/NguyenDT16} proposed a IM problem called \textit{Cost-aware Targeted Viral Marketing} (CTVM), which considers the heterogeneous costs of activating nodes and benefits of influenced nodes. The goal is to find the most cost-effective seed users who can influence the most relevant users to an advertisement. 
Guo \etal~\cite{DBLP:conf/cikm/GuoZZCG13} studied the problem of finding a small number of nodes that can maximize the impact on a given target user (\ie local optima) in a network. The proposed method simulates the objective function by a random function with a low variance guarantee under the IC. \eat{Moreover, in order to be applicable to the online scenario, a scalable approximation algorithm is proposed. The algorithm works by constructing a local cascade structure community that consists of the shortest paths between each node and the target node in the graph.}

Despite the progress made by the aforementioned methods, efficiency remains a challenge for TIM. In the topic-aware influence propagation model, each edge in a social network depends on a topic, which means that different queries correspond to different topic distributions. This makes query evaluation expensive in practice due to the large number of topics. Furthermore, existing methods formalize topics into probability distributions in potential numerical spaces. This makes it difficult for end users to specify such distributions.

\vspace{-2ex}\subsection{Location-aware IM}
%
Due to the proliferation of smart devices, location-based social networks (LBSN) have provided a new channel for us to bridge online and offline user activities. By taking into account location information, there have been several studies focusing on \textit{location-aware IM} problem, which can be broadly classified into two types: (a) Given a query region, find a seed set in the given region to maximize influence propagation where each user is associated with a geographic location. (b) Given a query location, find a seed set that maximizes the weighted aggregated influence over other users, where each user is assigned a weight inversely proportional to the distance between the user and the query location. \eat{Fig.~\ref{fig:fig_loc} depicts an example of a query region and query location.} The challenge for location-aware IM is efficiency as the solution has to be carried out based on the query location/region, which is submitted in real time.

\btitle{\textit{Type} (a).} According to \cite{DBLP:conf/sigmod/LiCFTL14}, every user in a social network has a location in 2D space. Given a query region, the authors attempt to select a group of users that will have the maximum influence on users in the query region. The influence spread in a user's area is calculated by extending the \textsf{PMIA} model (introduced in Section~\ref{ssec5.2}), where two filtering strategies are proposed. The first one is based on expansion to estimate the upper limit of user influence. The second strategy pre-calculates users of a small area and their influence by utilizing spatial indices. This framework assumes each user is static. Zhou \etal~\cite{DBLP:conf/cikm/ZhouCLXZL15} studied the influence maximization by considering mobility of users. This method can evaluate the possibility of users adopting a certain product according to the location of users and products, and then estimate the online influence to be converted into the possibility of physical store sales. Notably, these works~\cite{DBLP:conf/sigmod/LiCFTL14,DBLP:conf/cikm/ZhouCLXZL15,DBLP:conf/kdd/ZhuPCZZ15} ignore the impact of distance between an arbitrary user and the queried location/area on the final results.

\btitle{\textit{Type} (b).} In order to overcome the above limitation, Wang \etal \cite{DBLP:conf/icde/WangZZL16,DBLP:journals/tkde/WangZZL17} proposed \textit{distance-aware influence maximization} (DAIM) problem. The authors presented two methods, \textsf{MIA-DA} and \textsf{RIS-DA}, under the IC model. \textsf{MIA-DA} extends the tree model of~\cite{DBLP:conf/kdd/ChenWW10} and can achieve an approximate ratio of $1-1/e$. On the other hand, \textsf{RIS-DA} extends the RIS model and can achieve a ratio of $1-1/e-\varepsilon$ with a probability of at least $1-\delta$. The main steps are as follows. Firstly, it uniformly samples a set of locations randomly from user distribution in 2D space and pre-computes the range of influence relative to the sampled positions. Then, by evaluating the relationship between the query location and the sampled one, it can generate the influence spread of any query location. In \textsf{MIA-DA}, the space is divided into multiple units of identical size and the center of each unit is selected as the sampled location. Afterwards, the whole space is divided into Voronoi cells accordingly. This ensures that the nearest location with respect to a query can be easily acquired from the sampled ones. However, both methods need to estimate the size of the network samples required for any potential query, which is extremely time consuming. Recently, Zhong \etal \cite{DBLP:conf/dasfaa/0002ZZLQ18} advocated that the real query position is not any point in space but should be close enough to some users, \ie the query position is always consistent with the user distribution. Based on that, they proposed a new problem of locating new facilities among $n$ given points. By assuming that the potential query location does not exceed a certain distance from a user, the problem is simplified to the \textit{$l$-center problem}~\cite{DBLP:journals/ipl/FowlerPT81}. A heuristic algorithm is presented to return $l$ points as the final sample positions. Song \etal \cite{DBLP:conf/cikm/SongHL16} simultaneously consider the factors of time and geographic constraints by introducing a login probability for each user, and propose a sampling-based method called \textsf{Target-IM}. This method introduces \textit{weighted reverse reachable} \textit{(WRR) trees} and then greedily selects $k$ nodes that can cover the largest number of WRR trees.

As can be seen from the above, the location-aware IM problem is defined in a variety of ways. Despite these variations, only a few have considered bridging the online behavior and offline locations, which is of much significance for understanding human behavior and community activities.

\vspace{-2ex}\subsection{Social psychology-aware IM}
As discussed in~\cite{DBLP:conf/kdd/KempeKT03}, information diffusion process in social networks relies on the ``word-of-mouth effect'' between individuals. In particular, the ``word-of-mouth effect'' between a pair of individuals is influenced by human traits. In fact, there have been plenty of research in social psychology to investigate these traits~\cite{citeulike:436278}. Among these characteristics, \textit{conformity} is an important one that plays pivotal role in social influence ~\cite{CG03,EG99}. It refers to the inclination of an individual to be influenced by others by yielding to perceived group pressure and copying the behavior and beliefs of others. It is well known that humans will readily conform to the wishes or beliefs of others~\cite{CG03}. It was perhaps a surprise when~\cite{Asch55} found that people will do this even in cases where they can obviously determine that others are incorrect. The effect of conformity has been also justified in neuroscience study~\cite{ESDD11,KMSF11}.
However, majority of online information diffusion and IM studies fail to consider such social psychology factors. In particular, the propagation probability from $u$ to $v$ typically captures how it can be affected by the influence of $u$ but not conformity of $v$. In recent times, this has led to research that employ these social psychology factors in order to estimate the actual influence of nodes more accurately.

Li \etal~\cite{DBLP:conf/edbt/LiBS13} are the first to incorporate conformity in the IM problem. They proposed a \textit{conformity-aware cascade} ($C^2$) model, which leverages on the interplay between influence and conformity in obtaining the propagation probabilities of nodes from underlying data for estimating influence spreads. \eat{Under the $C^2$ model, each node in a network is associated with a pair of scores indicating its influence and conformity, respectively. According to $C^2$, the propagation probability from $u$ to $v$ not only depends on the influence of $u$, but also on the conformity of $v$. They presented a solution towards the IM problem under $C^2$ following the snapshot-based sampling approach.} This effort is extended in~\cite{DBLP:journals/vldb/0005BSC15} as follows. Firstly, it considers \textit{context-specific} influence and conformity of nodes and present a new cascade model, $C^3$, which incorporates topic-aware influence and conformity into $C^2$. Secondly, it presents an end-to-end framework for addressing IM under $C^3$ from UGCs in a given social network. \eat{In particular, in order to evaluate the context-aware conformity and influence scores for each individual} It performs the following steps: (a) extract the topic information via keywords and cluster the UGCs accordingly; (b) evaluate the sentiment expressed in UGCs by utilizing the technique in~\cite{Jo:2011:ASU:1935826.1935932}; (c) for each topic cluster, compute the influence and conformity scores for each individual according to the algorithm introduced in~\cite{LBS11}.

More recently, \cite{DBLP:conf/wcsp/LiGFTQZ18} further classifies conformity into \textit{friend}  and \textit{group} conformity. The former refers to one's inclination to conform to her friends' whereas the latter refers to one's inclination to behave in the same way to the group she belongs to. \eat{They advocate that the friend conformity can be attributed to the similarity between a pair of friends' profile, while the group conformity of a node $u$ can be defined as the similarity between the profile of $u$ and that of the group $u$ belongs to. A node $u$'s profile is defined as a series of frequent topics that appeared in $u$'s UGCs, while a group profile is defined as top-$\ell$ frequent topics that appeared in UGCs of all members.} Based on the two conformity behaviors, they propose a new cascade model where the probability of $u$ influencing $v$ is the product of their friend conformity and the group conformity of $u$. Accordingly, a \textit{group-based} influence maximization problem to select $k$ seeds to maximize the influence spread under the conformity-aware diffusion model is presented.

It is well-known that social psychology plays a pivotal role in governing social behavior of humans. Social psychology-aware IM is among the most recent addition to the arsenal of IM research compared to other extended IM problems (Fig.~\ref{fig:Extended_IMwork}). There are many theories in social psychology including conformity. All the aforementioned work, however, focus on conformity. Several other theories that govern human behavior such as confirmation bias, self-affirmation, are yet to be explored in the context of online information diffusion and influence maximization.

\vspace{-2ex}\subsection{Time-aware IM}\label{ssec64}
Some marketing strategies may be time-sensitive as a stakeholder may often give a deadline for maximizing influence spread of a particular product. Furthermore, influence propagation process itself takes time and the ability of information to spread typically weakens over time. Consequently, it is necessary to incorporate the time factor in IM for certain applications. This type of IM problem is referred to as \textit{Time-Aware Influence Maximization}.
Specifically, the influence propagation in time-aware IM has to consider two aspects, time decay and time delay propagation. The phenomenon of time decay captures the fact that information transmitted is time-sensitive, it will not propagate permanently, and the influence of information will decay over time. Time delay encapsulates the phenomenon that information can spread from one person to another but there may be a certain delay. Hence the speed of propagation varies during the diffusion process.

Saito \etal \cite{DBLP:conf/acml/SaitoKOM09} proposed an information diffusion model by incorporating continuous time delay based on the IC model, namely \textit{continuous time delay independent cascade } (\textsf{CTIC}) model. The model achieves a continuous time delay by introducing a time-delay parameter $r$ and a diffusion parameter $k$ for each edge $e = (u, v)$. Rodriguez \etal \cite{DBLP:conf/icml/Gomez-RodriguezBS11} introduced the \textit{continuous-time generative} model for cascade data generation in social networks by considering time delay propagation. They modeled the diffusion as a spatially discrete network of continuous, conditionally independent time process occurring at different rates, by associating each edge with a transfer function (density over time). Notably, these efforts do not consider the influence maximization under time constraints.

Chen \etal \cite{DBLP:conf/aaai/ChenLZ12} maximized the influence spread over a given period of time and proposed the \textit{Independent Cascade with Meeting} \textit{events} (\textsf{IC-M}) model to capture the time delay of propagation by assuming that influence delays follow the geometric distribution. Under the \textsf{IC-M} model, two heuristic algorithms based on the concept of Maximum Impact Tree Structure (MIA) \cite{DBLP:conf/kdd/ChenWW10} are proposed. The first is based on a dynamic programming process that computes the precise effects in the tree structure. The second converts the problem to fit the classical IC model such that any existing heuristic approach for classical IM can be applied. Subsequently, Liu \etal \cite{DBLP:conf/icdm/LiuCXZ12} proposed the time constrained influence maximization problem under a delay-aware independent cascade captured by the \textit{Latency Aware Independent Cascade} (\textsf{LAIC}) model, under which a time step based simulation algorithm was designed to estimate the information diffusion of seed sets under time constraints. Different from \cite{DBLP:conf/aaai/ChenLZ12}, this work is also applicable when other distributions are used in the model. The two models, \textsf{LAIC} and \textsf{CTIC}, are essentially the same when node $u$ is activated at step $t$, \ie it activates its currently inactive neighbor $v$ with probability $p$ in step $t + \delta_t$, where $\delta_t$ is the delay and is randomly derived from the delay distribution. In the fully continuous time-diffusion model introduced in \cite{DBLP:conf/icml/Gomez-RodriguezBS11}, Gomez-Rodriguez \etal \cite{DBLP:conf/icml/Gomez-RodriguezS12} used the lazy evaluation strategy to reduce the number of evaluation of marginal returns.

The above methods do not provide any approximation guarantee. Du \etal \cite{DBLP:conf/nips/DuSGZ13} proposed a randomized algorithm for the influence estimation in continuous time diffusion networks. The main idea is to simplify the problem into the neighborhood estimation problem in the graph from the perspective of graphical model inference. This method is guaranteed to find the influence of seed set with at least $(1-1/e) OPT-2k\varepsilon$. Ohsaka \etal \cite{DBLP:conf/pkdd/OhsakaYKK16} proposed a new diffusion model incorporating the time decay probability and time delay, namely \textit{Time-varying IC} model. Under this model, an optimal solution with an approximation ratio $(1-1/e-\varepsilon)$ can be obtained by following a standard hill-climbing strategy. It is worth noting that all previous propagation models were extended based on the IC model. In  \cite{DBLP:conf/pkdd/OhsakaYKK16}, a time-aware LT model is also proposed. Huang \etal \cite{DBLP:conf/icde/HuangBCZ19} used Monte Carlo simulation to randomly generate subgraphs from snapshots of social networks at different times, and employ two graphics compression technologies to alleviate the problem of high memory usage.

The above methods mainly work by adding time factors to the propagation model. In addition, most of these models are extended based on the IC model. There is scant efforts to extend other diffusion models such as LT. One possible future direction is to consider the phenomenon that each person's rate of information propagation is different depending on both time and topic.

\vspace{-2ex}\subsection{Dynamic IM}\label{ssec63}
The majority of research in IM assume that the target network is static. However, in reality, social networks evolve with time. As new interactions emerge and old edges become obsolete, the topology of a network changes rapidly. \eat{For instance, on Facebook, when a user follows/unfollows another user, the link appears or disappears, respectively.} In addition, the intensity of influence is constantly changing. Therefore, it is necessary to select the seed set in real-time by considering the evolution of the network structure. We refer to this as \textit{dynamic} IM problem.

Aggarwal \etal \cite{DBLP:conf/sdm/AggarwalLY12} are the first to introduce the dynamic model for information flow in social networks. They combined the greedy approach and forward temporal analysis to determine the most influential points in a network. Specifically, this approach finds a seed set at time $t$ under the condition that the network evolves within a given interval $[t, t + h]$. It aims to maximize the influence at $t + h$. Zhuang \etal~\cite{DBLP:conf/icdm/ZhuangSTZS13} proposed \textit{Maximum Gap Probing} (\textsf{MaxG}), which attempts to observe the network changes by periodically detecting the connection of a small number of nodes to minimize the information diffusion gap between the observed and the actual network. \eat{This is achieved by processing an observed network at each timestamp $t$ to handle changes to it.} However, these methods do not consider the maximization of influence under diffusion models such as IC/LT. Additionally, all are heuristic algorithms without any performance guarantee.

Chen \etal~\cite{DBLP:conf/sdm/ChenSHX15} consider modeling dynamic social networks as a series of snapshot graphs. They extended IM to the \textit{Influential Node Tracking} (INT) problem, which focused on tracking a set of influencing nodes that maximize the influence as a network evolve. They proposed the \textit{Upper Bound Interchange Greedy} (\textsf{UBI}) algorithm, which uses the smoothness of evolution of the network structure to track the seed set found in the preceding snapshot and changes the nodes in the preceding seed set one by one to discover more influential nodes incrementally. The algorithm can achieve a $1/2$ approximation ratio over the result quality. However, \textsf{UBI} is sensitive to the size of the seed set. As the size increases, its performance and solution quality decrease. Unlike \textsf{UBI}, the method proposed by Ohsaka \etal \cite{DBLP:journals/pvldb/OhsakaAYK16} can incorporate updates, including addition and removal of vertices and edges, and updates of propagation probabilities. They achieved that by maintaining a real-time dynamic index of a series of graph sketches following RIS strategy. By presenting a specific updating scheme for the index, they achieve a $(1-1 /e-\varepsilon)$ approximation ratio. \eat{In particular, the graph is updated by updating a sketch.
First, it needs to build an index on the RR set based on the initial graph. As long as the total weight of the current index is less than $W= \Theta(\varepsilon^{-3}(n+m)\log n)$, it repeatedly creates new sketches and adds them to the index. In particular, each sketch is associated with a weight, which equals to the space required to store the sketch, \ie the number of nodes plus the number of outlinks inside the sketch. When the graph is updated, a reverse BFS is performed on the target vertex in the index based on the original sketch. Whenever the accumulated weight for all the sketches insides the index is greater than or equal to $W$, it needs to delete the last sketch of the index.} However, this method can only handle updates to hundreds of nodes/edges per second due to the high maintenance cost. Yang \etal \cite{DBLP:conf/cikm/0001WJPC19} proposed a method to track the top $k$ influential users by controlling the relative error of the results. It maintains a proper number of RR sets for approximating users' influence spreads.

These dynamic IM solutions either fails to provide a theoretical guarantee of the result quality \cite{DBLP:conf/sdm/AggarwalLY12,DBLP:conf/icdm/ZhuangSTZS13} or provide a guarantee at the expense of high updating overhead \cite{DBLP:conf/sdm/ChenSHX15,DBLP:journals/pvldb/OhsakaAYK16}. Wang \etal \cite{DBLP:journals/pvldb/WangFLT17,DBLP:journals/tois/WangLFT18} defined a novel IM query model called \textit{Stream Influence Maximization} (\textsf{SIM}) to track influential users in real time. \textsf{SIM} mainly uses the sliding window model to find $k$ users with the greatest influence in the current window by considering the most recent $N$ actions. This process handles changes in the network structure by maintaining a series of influential checkpoints. Wang \etal \cite{DBLP:journals/geoinformatica/WangLYH19} proposed a regularized learning framework for modeling topic-aware influence propagation in dynamic network structures. However, instead of directly addressing the IM problem, they studied how the topical content spreads in dynamic relational networks. \eat{The framework mainly finds influential users through content and structural information and use them as a potential feature of observation data.} It constructs a Markov chain on the feature representation to model the dynamic evolution. Unlike the traditional sliding window model, Zhao \etal \cite{DBLP:conf/icde/ZhaoSWLZ19} proposed a time-decaying dynamic interaction network (TDN) model to evaluate the node interaction stream, which can smoothly discard outdated user interactions.

Although these efforts have taken into account the evolution of networks, unfortunately the influence of an arbitrary seed (set) is estimated on the basis that the network is static. For instance, even if we sample a snapshot of a dynamic network at time $t$, referred to as $G_t$, and obtain a seed $v$ with the maximal influence over $G_t$, $v$ may not eventually influence the most nodes in practice as the network will evolve during influence propagation from $v$. Therefore, how to evaluate the influence of seeds by taking into account the evolution during influence propagation remains an open problem.

\vspace{-2ex}\subsection{Competitive IM}
The traditional IM problem aims to maximize influence by assuming that only one organization is involved for choosing $k$ influential users. However, in practice, viral marketing is always performed in competitive scenarios (\ie marketing makes little sense in monopolized scenario). Hence it is important to consider competitive products and their corresponding marketing strategies during influence maximization. \eat{For instance, in the mobile phone market, there are multiple competitors (\eg Apple, Huawei, Samsung) trying to maximize influence of their products.} The goal of \textit{Competitive Influence Maximization} problem is to find a group of seed users to disseminate information in order to maximize the number of affected nodes in a competitive environment where other competitors also engage in information broadcasting activities. The problem assumes that multiple products spread across social networks and consumers may use products exclusively. This problem is particular pertinent for a company with a smaller marketing budget to effectively penetrate a highly competitive market.

Bharathi \etal \cite{DBLP:conf/wine/BharathiKS07} and Carnes \etal \cite{DBLP:conf/ACMicec/CarnesNWZ07} are the first to consider competition in the IM problem. They assumed that the seeds of an opponent's initial choice are known in advance. From the perspective of game theory, the IM problem is transformed into a game with the aim to find the best response to other competitors in the game, \ie find the Nash Equilibrium of the game. They prove that maximizing the influence in this environment is also NP-hard and submodular. Therefore, they proposed a pair of hill-climbing algorithms to maximize influence based on the prior knowledge of choices made by the competitors. Both methods are based on extensions of the IC model. Borodin \etal \cite{DBLP:conf/wine/BorodinFO10} presented  several extensions of the LT model in a competitive environment. However, influence maximization under the proposed model is not submodular. Their solution does not provide any theoretical guarantee over the results. Budak \etal \cite{DBLP:conf/www/BudakAA11} and He \etal \cite{DBLP:conf/sdm/HeSCJ12} observed that different information has different acceptance rates. An entity can strategically select some seed nodes that can initiate its own influence transmission to prevent the influence of its competitors from spreading as much as possible. Thus, they defined the \textit{influence blocking maximization} (IBM) problem as choosing positive seed nodes in a social network in order to minimize the spread of negative effects or to maximize the blocking effect on negative effects. Specifically, \cite{DBLP:conf/www/BudakAA11} is based on an extension of the IC model whereas \cite{DBLP:conf/sdm/HeSCJ12} is based on an extension of the LT model. In addition, \cite{DBLP:conf/sdm/HeSCJ12} proved that the objective function of influence blocking maximization under this model is submodular, so the greedy algorithm can achieve the approximation ratio of $1-1/e$.

The above methods aim to select a group of users to maximize the expected propagation under the assumption that the users selected by a competitor are known. That is, the strategy (\eg the algorithm used to select seeds) employed by an opponent is known. However, this assumption is too strict and unrealistic. An opponent's strategy is a business secret and it is not possible for a company to know this unless there is commercial espionage or information leakage.

Several works abandon such restrictive assumption and consider more realistic scenarios. Clark \etal \cite{DBLP:conf/gamesec/ClarkP11} assumed that every participant can select a seed incrementally and iteratively, like playing chess. Lin \etal \cite{DBLP:conf/kdd/LinLC15} proposed a learning-based framework to address the \textit{multi-round} competitive influence maximization problem on social networks. The opponent's strategy in this question can be known or unknown. \eat{The company needs to make sequential decisions in multiple rounds to maximize the expected influence in the long run.} It trains the model to learn a strategy to defeat the opponent's strategy gradually. \eat{If the opponent's strategy is unknown, then seeking a Nash Equilibrium for the process is the best strategy a rational company should take.}  Li \etal \cite{DBLP:conf/sigmod/LiBCGM15} also modeled the IM problem in a competitive network as a game using each group as a competitor and rewarding the expected influence under a strategy. They redefined real-time tasks in the network to the best real-time strategy choices in the game. Then whether there is a Nash Equilibrium is explored and a strategy of Nash Equilibrium in a competitive network with $n$ participants and $z$ strategies is presented.

Almost all models in the aforementioned studies assume that the propagating entity is in a purely competitive state. Lu \etal \cite{DBLP:journals/pvldb/LuCL15} and Ou \etal \cite{DBLP:conf/cikm/OuCC16} consider a new competitive IM problem in which the diffusion model captures both competitive and complementary states. A process of reducing or increasing the likelihood of using a second product after using the first is adopted to model a competitive and complementary relationship. In terms of modeling, \cite{DBLP:journals/pvldb/LuCL15} extended the IC model to the \textit{comparative IC} (\textsf{Com-IC}) model and added additional mechanisms to model products that are reconsidered with other products. In \cite{DBLP:conf/cikm/OuCC16}, the LT model is extended, which takes into account the collaborative impact of neighbors on each possible product. Recently, a new competition problem has been proposed \cite{DBLP:conf/infocom/ZhuLZ16,DBLP:journals/www/YanZLY19} to select seeds at minimal cost to compete with others in social networks and to spread the information satisfying a preset threshold. In addition, Banerjee \etal \cite{DBLP:journals/corr/abs-2012-03354} studied the competitive social welfare maximization problem under the utility-based independent cascade model(UIC). The node in this problem can adopt multiple items, and the goal is to maximize the overall welfare of all items. In the case of arbitrary interactions between items, the node uses chosen value functions to model the adoption of multiple items. The problem is neither monotone nor submodular. They proposed the SeqGRD solution, which can provide $\frac{u_{\min }}{u_{\max }}\left(1-\frac{1}{e}-\epsilon\right)$- approximation guarantee, where $ u_{\min }$ is the minimum expected utility among all individual items, $u_{\max }$ is the expected maximum utility among all item bundles.

In summary, two categories of efforts aim to address the Competitive Influence Maximization problem. One is to extend existing information diffusion models to incorporate competitive influence. The other is to represent different competitors as players to model the problem as a game from the perspective of game theory. The main goal is to find a state in which any group in the competition cannot influence more users by changing its strategy, \ie the Nash Equilibrium.

\vspace{-2ex}\subsection{Budgeted IM}
In the field of online advertising services, there definitely exists some budget before the diffusion of an advertisement. Generally, given a social network and a fixed budget, there may be different costs for activating nodes in the network.  \textit{Budgeted Influence Maximization} (BIM) is to select a set of seed nodes to spread information, in order to maximize the total number of nodes influenced in the social network, subject that the total cost does not exceed the budget.

Nguyen \etal \cite{DBLP:journals/jsac/NguyenZ13} proposed two heuristic methods to construct directed acyclic graphs from a general graph that captures the bulk of influence spread. They consider the influence diffusion calculation problem on the directed acyclic graph as an example of belief propagation on the Bayesian network. They provide a solution with $(1-1/\sqrt{e})$ approximation ratio. Han \etal \cite{DBLP:conf/pakdd/HanZHS14} proposes three heuristic strategies to solve the problem. The heuristics they employ to select seeds are, most influential node, most cost-effective node, and hybrid selection from both. But the above heuristics algorithm provides no guarantee on the returned answer.

Different from the heuristic algorithm, Bian \etal \cite{DBLP:journals/pvldb/BianGWY20} proposed a new method based on reverse sampling to solve the BIM problem. The algorithm can find the seed set with the highest expected influence under budget constraints, and terminates after reaching $1-1/{e^\beta-\varepsilon}$-approximate guarantee where  $1-1/{e^\beta} = (1-\beta)(1-1/e)$. The previous BIM algorithm mainly considered non-adaptive settings, that is, selecting all seed nodes in one batch without knowing how to affect other users. Recently, Huang \etal \cite{DBLP:conf/icde/Huang0XSL20} proposed the adaptive target profit maximization problem, and studied the problem in both oracle model (the profit of any node set can be accessed in $O(1)$ time) and noise model (only expected spread can be estimated). In addition, they proposed a pair of algorithms for both models, respectively, mainly using the actual influence knowledge in previous batches to guide the selection of subsequent batches.
\vspace{-2ex}\subsection{Adaptive IM}

Existing IM problem utilizes a diffusion model as an input to describe how information is propagated through a network where the influence probability is known. In addition, the seed set $S$ needs to be determined before the influence propagation process. Unfortunately, the basic diffusion model and its parameters are difficult to obtain in practical applications. Therefore, if the selection of seed nodes is entirely driven by a diffusion model and influence probability distribution, this may lead to a serious overestimation of the actual spread of the selected seed nodes. To address this problem, \textit{Adaptive Influence Maximization} (AIM) problem was proposed. Hereby \textit{adaptive} means that the seed candidate can be selected in multiple rounds during the propagation of influence. An AIM algorithm will be able to observe the propagation results of the preceding round and use this knowledge to select the candidate seeds for the next round. Formally, this problem is defined as:

\vspace{-1ex}\begin{definition}
	{\em Given a social network $G = (V, E)$, integer $k$, the \textbf{adaptive influence maximization} (AIM) problem aims to select $k$ nodes as the seed set $S (S \subseteq V )$ through an adaptive policy $\pi^{*}$, such that:
	\begin{equation}
    \pi^{*} \in \arg \max _{\pi} f_{a v g} \triangleq \mathbb{E}_{\Phi}[f(E(\pi, \Phi), \Phi)] \quad \text { s.t. } \quad \forall \phi, |E(\pi, \phi)| \leq k
	\end{equation}
where $\phi$ is the realization of the influence graph (true world), $E(\pi, \phi)$ represents the seed nodes that have been selected following policy $\pi$ under realization $\phi$.}
\vspace{-1ex}\end{definition}

Depending on whether the information diffusion model is known or not, existing AIM solutions can be categorized into two groups: Rows 1-7 of Table~\ref{table:adaptive IM} (with known diffusion model); Rows 8-13 of Table~\ref{table:adaptive IM} (diffusion model unknown). We shall elaborate on both types in turn.

\begin{table}[t]
	\centering
	\vspace{0ex}\caption{Adaptive Influence Maximization algorithms.}\vspace{-1ex}
	\resizebox{\textwidth}{!}{
	\begin{tabular}{ccccc}
		\toprule  
		IM Methods& Feedback Type& Information Diffusion model& Approximation Bound& Adaptive Submodular \\
		\midrule  
		Golovin \etal \cite{DBLP:journals/jair/GolovinK11}& Full Feedback& IC Model& $1-1/e$& Yes \\
		Vaswani \etal \cite{DBLP:journals/corr/VaswaniL16}& Full Feedback& IC Model& $1-e^{-\frac{1}{\gamma}} \quad and \quad \gamma=\left(\frac{e}{e-1}\right)^{2}$ & Yes \\
		Han \etal \cite{DBLP:journals/pvldb/HanHXTST18}& Full Feedback& IC Model& $ \left\{\begin{array}{ll}{1-\exp (\xi-1),} & {\text { if } b=1} \\ {1-\exp \left(\xi-1+\frac{1}{e}\right),} & {\text { otherwise }}\end{array}\right.$& Yes \\
		Yuan \etal \cite{DBLP:conf/ijcai/YuanT17}& Partial Feedback& IC Model&$\alpha\left(1-e^{-\frac{1}{\alpha}}\right)\quad \alpha=1 (Full Feedback)$ & No \\
		Salha \etal \cite{DBLP:conf/asunam/SalhaTV18}& Myopic Feedback & Modified IC Model& $1-1/e$& No \\
		Peng \etal \cite{ DBLP:conf/nips/PengC19}& Myopic Feedback& Modified IC Model& $1/4(1-1/e)$& No \\
		Sun \etal \cite{DBLP:conf/kdd/SunHYC18}& $\quad$& Multi-Round IC Model& $ \left\{\begin{array}{ll}{\frac{1}{2}-\varepsilon,} & CR-Greedy \\ {0.46-\varepsilon,} & WR-Greedy\end{array}\right.$& No \\
		\midrule  
		Chen \etal \cite{DBLP:conf/icml/ChenWY13}& Edge Feedback& IC Model& $1-1 / e-\varepsilon, 1-1 /|E|$& $\quad$ \\
		Lei \etal \cite{DBLP:conf/kdd/LeiMMCS15}& Edge Feedback& WC Model& $\quad$& $\quad$ \\
		Wen \etal \cite{DBLP:conf/nips/WenKVV17}& Edge Feedback& IC Model& $\quad$& $\quad$ \\
		Vaswani \etal \cite{DBLP:journals/corr/VaswaniKWGLS17}& Node Feedback& IC and LT Model& $1-1/e$& $\quad$ \\
		Vaswani \etal \cite{DBLP:journals/corr/VaswaniL15}& Edge and Node Feedback& IC Model& $1-1 / e-\varepsilon, 1-1 /|E|$& $\quad$ \\
		Lagr\'{e}e \etal \cite{DBLP:conf/icdm/LagreeCCM17,DBLP:journals/tkdd/LagreeCCM19}& Node Feedback& TV and IC Model& $\quad$& $\quad$ \\
		\bottomrule 
	\end{tabular}
}
	\label{table:adaptive IM}\vspace{-1ex}
\end{table}
\vspace{0ex}\subsubsection{Adaptive IM with Known Information Diffusion Model}
Golovin \etal \cite{DBLP:journals/jair/GolovinK11} is the first to extend the definition of submodularity and monotonicity to the adaptive setting. In this setting, a batch of nodes is seeded at different time intervals. When a batch is seeded, it eventually spreads according to the classical IC model. Then the algorithm will select the next batch of seeds based on the previously observed cascade. They proved that if the objective function satisfies that the marginal gain of each element for each possible realization does not increase along with the size of the group, it is adaptively monotone and submodular. They assume an \textit{edge level feedback} mechanism under the IC model and show that the expected spread is adaptively monotone and submodular. Consequently, a hill-climbing strategy guarantees an approximation ratio. The challenges in this problem setting are how many seeds should be selected for each batch and how long should it wait between successive seeds. If the waiting time is long enough, selecting one seed at a time and waiting for the diffusion to complete before the next selection should result in maximum propagation \cite{DBLP:journals/jair/GolovinK11}. In reality, there may not be sufficient time to complete a diffusion. Vaswani \etal \cite{DBLP:journals/corr/VaswaniL16} proposed to select $k$ seed nodes in $r$ rounds, and select $b$ nodes in each batch $(b = k/r)$. After selecting each batch to observe the eventual influence, the ultimate goal is to select $r$ seed sets to maximize the expected influence distributed over the possible worlds. However, \cite{DBLP:journals/corr/VaswaniL16} needs to estimate the expected influence of all seed sets, which will lead to excessive computational overhead. In addition, since the method used to select each seed node is invalid under the adaptive setting, this solution does not provide any non-trivial approximation guarantee. Han \etal \cite{DBLP:journals/pvldb/HanHXTST18} proposed the \textsf{AdaptGreedy} framework, which uses existing non-adaptive IM methods as building blocks. This method has a larger error in estimation of expected spread than \cite{DBLP:journals/corr/VaswaniL16}. \eat{In addition, they proposed a non-adaptive \textsf{EPIC} algorithm based on the RR set method, which starts from a small number of RR sets and iteratively increases the number of RR sets until a satisfactory expected spread is obtained.} Recently, Tang \etal \cite{DBLP:conf/sigmod/TangHXLTSL19} proposed a sampling method based on the concept of multi-root reverse reachable (mRR) sets. On the basis of this sampling method, an algorithm is proposed to maximize truncated influence spread with a provable approximation guarantee of $(1-1 / e)(1-\varepsilon)$.  \cite{DBLP:conf/sigmod/TangHXLTSL19} adaptively selects the node with the maximum expected truncated influence spread in each round of seed selection, ignoring the other influence diffusion that may lead to incorrect selection of seed nodes.

Yuan \etal \cite{DBLP:conf/ijcai/YuanT17} introduced the \textit{Partial Feedback} model that captures the trade-off between latency and performance. It performs the selection process in a sequential manner, where each round of decisions depends on the current observation of the state of the network and the remaining budget. \eat{When certain conditions are met, the next seed is selected, otherwise, wait for a time slot and update the network realization. In addition, they proposed a $(\alpha,\beta)$-\textit{greedy} strategy, which guarantees a constant approximation ratio under this model.} Sun \etal \cite{DBLP:conf/kdd/SunHYC18} proposed a \textit{Multi-Round Triggering} (\textsf{MRT}) information diffusion model based on the Triggering (IC) model~\cite{DBLP:conf/kdd/KempeKT03} to solve the problem of adaptive multi-round IM. The \textsf{MRT} information diffusion model consists of $T$ independent rounds of diffusions. During each round, it starts with a separate seed set and performs the basic trigger diffusion independently. Salha \etal \cite{DBLP:conf/asunam/SalhaTV18} introduced the \textit{Myopic Feedback} model where after a seed node is activated at time $t$, the state of its neighbors can be observed only at time $t + 1$. They modified the traditional IC model by enabling each active node to have many opportunities to influence its inactive neighbors. In addition, a new utility function is introduced that takes into account the cumulative number of active nodes over time. \eat{The utility function is adaptively monotone and submodular only in the modified IC model.} Peng \etal \cite{ DBLP:conf/nips/PengC19} is also a study on the adaptive influence maximization with \textit{Myopic Feedback} under the IC model. They proved that the upper and lower bounds of the adaptive gap are constant. 

\vspace{-1ex}\subsubsection{Adaptive IM with Unknown Information Diffusion Model}
Since it may be unrealistic to assume that the parameters of a diffusion model and influence probabilities are known \textit{apriori}, some scholars have recently leveraged on the multi-armed bandits (\textsf{MAB}) technique to learn potential diffusion parameters during multiple rounds of seeds selection. In the traditional MAB framework, there are $m$ arms and each arm has a random variable that represents the reward. The boundary of the reward is within $[0, 1]$, and remains independently and identically distributed according to some unknown distribution. In each round $s$, the bandit plays an arm and obtains a corresponding reward by sampling the reward distribution for that arm. The game continues for a fixed $T$ rounds. The goal is to minimize the regrets caused by playing suboptimal arms in each round. For each round of choice, the main problem is modeled as \textit{Exploitation-Exploration}, where \textit{Exploitation} refers to the process of pulling the arm with the highest expected reward and \textit{Exploration} refers to sampling arms to get more information about their expected benefits.

Compared to the \textsf{MAB} problem, the IM problem is equivalent to a combinatorial nature between multiple choices with a non-linear reward function. Chen \etal~\cite{DBLP:conf/icml/ChenWY13} studied the IM semi-bandit with edge-level feedback under the IC model.
Herein, edge-level feedback refers to the following settings, during multiple rounds for the $k$ seeds selection, where the propagation probability via each edge is unknown, the seeds selection in consecutive rounds is adjusted accordingly by observing all eventual edge activations since the last round of selection. Based on that, \textsf{CMAB} is proposed, where each edge is regarded as a branch, and the super branch is a set of edges output from at most $k$ nodes.
The framework uses this information from the previous round to decide which super arm to play in the next round, while learning the activation probability of the edge. In addition, the framework assumes that the oracle provides an $(\alpha,\beta)$-approximation to the optimal solution. The oracle outputs with probability $\beta$ a super-arm such that it attains an $\alpha$ approximation to the optimal solution. The method in \cite{DBLP:conf/icml/ChenWY13} counts activated nodes multiple times, resulting in redundant activation and selection. Beside, each arm needs to be tested for feedback, which can result in a large activation budget. Lei \etal \cite{DBLP:conf/kdd/LeiMMCS15} proposed an online influence maximization (\textsf{OIM}) framework, which learns the influence probability online. They proposed a multiple-trial approach. Firstly, the uncertainty over the influence is captured as a Beta distribution. Then, $k$ seed nodes are selected from the graph to obtain the maximum return or by estimating the confidence interval of the influence probability to improve the understanding of the diffusion. \eat{From these seeds, information diffusion is started and the user feedback obtained is used to update the influence estimation.} Wen \etal \cite{DBLP:conf/nips/WenKVV17} proposed the \textsf{IMLinUCB} algorithm, which considers a linear generalization model across edges and prove regret bounds for the special case of forests under edge-level feedback.

All of the above methods study edge-level feedback. However, in many cases the success/failure of edge activation attempts cannot be observed. Unlike observing edge states, observing the state of each node is easier and more intuitive. Vaswani \etal \cite{DBLP:journals/corr/VaswaniL15} proposed a ``node-level'' feedback mechanism by assuming that we can only observe whether each node is active, rather than knowing who is affecting it. $\varepsilon$-greedy and Thompson sampling algorithms are exploited to obtain this feedback model where the learning agent observes the affected nodes but not the edges. However, no theoretical guarantee is provided. More recently, Vaswani \etal \cite{DBLP:journals/corr/VaswaniKWGLS17} proposed a \textit{pairwise-influence }semi-bandit feedback model and develop a \textsf{LinUCB-based bandit} algorithm. The pairwise-influence feedback observes each node activation along with the seed node responsible for it. A parametric method through pairwise reachability probabilities is proposed to this end.  \cite{DBLP:conf/icdm/LagreeCCM17,DBLP:journals/tkdd/LagreeCCM19} defined the \textit{online influence maximization with persistence} (OIMP) problem where \textit{persistence} means that a node can be activated multiple times in different experiments but rewards are only counted once. They proposed the \textsf{GTUCB} (Good-Turing Upper Confidence Bound) algorithm, which relies on the well-known statistical tool called Good-Turing estimator~\cite{Good53}. Without knowing the diffusion model or propagation probability via each edge, they estimate the candidates' \textit{missing mass} by Good-Turing estimator. The \textit{missing mass} of a node is the expected number of nodes remaining to be activated. \eat{Based on the estimated \textit{missing mass}, the best candidate seed for the current spread can be determined accordingly.}

Table~\ref{table:adaptive IM} summarizes the characteristics and differences of the aforementioned work on AIM. AIM algorithms have the potential to increase the influence spread of a seed set in a real network by attempting to capture the diffusion process in a realistic setting. However, it can be seen that this setting does not consider the ``waiting time'' required to observe the influence of seed nodes before selecting the next batch. Hence the total running time of AIM is expensive in practice. In addition, adaptive submodularity is difficult to maintain for different problems. It is worth exploring how much the adaptive strategy outperforms non-adaptive strategies under various feedback mechanisms, and how to better balance waiting time and influence spread.
Last but not the least, the practical applications of AIM in viral marketing scenarios is still an open problem.

\vspace{-2ex}\subsection{Miscellaneous IM works}
There are other works that do not fall into any of the above categories. Although there is currently only a small amount of work on each problem, they are still acknowledged.

Tang \etal \cite{DBLP:conf/asunam/TangLZCZ14} proposed diversified social influence maximization, while considering the degree of influence and the diversity of the infected population. They constructed a class of diversity measures to quantify the diversity of the infected population. Antonio \etal \cite{DBLP:journals/tkde/CalioIPT18, DBLP:journals/isci/CalioT21} proposed Diversity-sensitive Targeted Influence Maximization. Different from \cite{DBLP:conf/asunam/TangLZCZ14}, this problem does not assume prior knowledge about user attributes. \cite{DBLP:journals/isci/CalioT21} associates the users of the social network with the classification data set, where each tuple expresses the classification attributes of the user. By designing non-decreasing monotone and submodular functions for categorical diversity, the concept of seed diversity is integrated into the objective function of IM problem. \cite{DBLP:journals/tkde/CalioIPT18} only considers using topological information in social networks to model user diversity. They use the definition of local diversity and global diversity to propose two greedy algorithms, which search for the shortest path in the diffusion graph in a backward manner from the selected target set. 
Yang \etal \cite{DBLP:conf/sigmod/YangMPH16} proposed Continuous Influence Maximization, and solved this problem through a coordinate descent framework. This problem assumes that every user in the social network knows the discount-related purchase probability curve, which is the seed probability function, and the goal is to select a group of users to provide corresponding discounts to maximize product purchases under a predetermined budget. Gao \etal \cite{ DBLP:journals/tcs/GaoGYYWX20} proposed the interaction-aware influence maximization problem, which takes into account both the number of active users and the interaction between users. The problem is non-submodular, so they solve it by decomposing the objective function into the difference between two submodule functions.

\cite{DBLP:conf/ijcai/TsangWRTZ19, DBLP:conf/www/FarnadiBG20} studied fair resource allocation in influence maximization. They introduced the concepts of equality and equity. Equality refers to the fair distribution of seeds to groups and is proportional to the size of the group in the population. Equity aims to minimize unfair results through fair treatment between different groups proportional to the size of the network. \cite{DBLP:conf/ijcai/TsangWRTZ19} proved that the IM problem of introducing fairness is non-sub-modular, so they transformed the problem into a multi-objective sub-module optimization problem and provided methods for multi-objective sub-module optimization. \cite{DBLP:conf/www/FarnadiBG20} transformed the problem into Mixed Integer linear Programming (MIP) problems, which is solved by the MIP solver. The fairness measures are encoded as a linear expression and included in the objective function.
\cite{DBLP:journals/is/LiCDWSX20, DBLP:journals/tkdd/GuoW20} are proposed to solve the community aware influence maximization problem, which tries to Maximize the number and community diversity of activated nodes. \cite{DBLP:journals/is/LiCDWSX20} proposed a Community-aware Partial Shortest Path tree (CPSP-Tree) to estimate the influence of candidate nodes. \cite{DBLP:journals/tkdd/GuoW20} proposed a Sampling-based Discretized-CG, which combines continuous greedy with Reverse Influence Sampling to improve its efficiency.

\vspace{-1ex}\section{The Road Ahead}\label{sec7}
Since the elegant seminal work by Kempe \etal~\cite{DBLP:conf/kdd/KempeKT03}, the IM problem has been extensively studied. In the preceding sections, we have discussed in detail key research efforts in this arena. The IM research began with the classical IM problem and since the turn of the last decade it diversified to several extensions that incorporate various real-world issues associated with information propagation in a social network. In this section, we delineate key future research directions in this area.

\textit{Realistic information diffusion models.} The dynamics of spread of information in a social network is guided by an information diffusion model. Hence, information diffusion models play a pivotal role in the IM problem. Inaccurate or overly simplistic diffusion models may adversely impact the result quality of the influence spread of a seed set. As discussed in this survey, majority of IM research have focused on utilizing the simplistic IC and LT models. Some of the research have also extended these models to cater for specific real-world issues such as location, conformity, and time. However, to ensure that the diffusion models closely capture real-world behaviors of social users and information propagation, it is paramount to explore holistic models that incorporate various real-world properties such as topics, social psychology theories, network dynamics. Currently, there is no diffusion model that takes into account \textit{all} of these features. Furthermore, IC and LT models consider influence propagation in discrete time steps whereas influence propagation occurs in continuous time in reality. Hence, diffusion models need to be augmented to capture this reality. To this end, Hawkes process-based models provide us a good starting point~\cite{DBLP:conf/aistats/ZhouZS13,Sigmodchasis2020}.

\textit{Realistic influence maximization algorithms.} Majority of existing IM techniques assign an influence probability between a pair of users that is typically computed without leveraging on several real-world properties of these users. For example, in the weighted IC model its is computed solely based on the degree of a node. This has considerable impact on the downstream computation of influence spread and seed sets. What properties should a ``realistic'' IM technique incorporate in computing the influence probability? Any item or information that a marketeer aims to propagate is associated with a topic. Furthermore, a social network is always evolving in reality. Lastly, behaviors of social users are always governed by social psychology theories. Hence, we advocate that any ``realistic'' IM technique should go beyond the classical IM paradigm by incorporating these three features (topics, dynamics, social psychology). Other features such as location, time-awareness, and competitive environment are application-dependent. For instance, not all IM applications need to be location-aware. Hence, these features need to be added based on the nature of specific applications that an IM algorithm is deployed to.

\textit{Realistic investigation of influence spread quality.}  Since the inception of IM research, experimental studies to verify the result quality of a seed set generated by an IM technique have traditionally been undertaken through simulation. However, such experimental setup does not really throw light on whether a user in a spread is influenced in reality. To elaborate further, consider a marketeer who wishes to promote a mobile phone by giving some of them free to the seed set $S$ detected by an IM algorithm. Let us denote the influence spread of $S$ as discovered by the algorithm as $I$. How many of these $I$ users are influenced in reality and subsequently bought the phone (this is the key goal of the marketeer)? Without systematic comparison with such information, the quality of influence spread (even with a theoretical guarantee) is unknown in practice. To this end, it is important to explore additional data sources (\eg purchase data, customer reviews etc) to measure the real impact of the seed set detected by an IM technique. This will shed light to the IM community on whether the result quality as measured in a long stream of publications is consistent with real-world data that serves as indicators of ``being influence''. There is a lack of systematic efforts to address this challenging question. We believe a large-scale user behavioral study is necessary to comprehend the real-world impact of existing IM techniques.

\vspace{-1ex}\section{Conclusions}\label{sec8}
In this paper, we present a comprehensive survey of the influence maximization research landscape. We review the information diffusion models and analyze a variety of algorithms for the classical IM problem. We propose a taxonomy for potential readers to understand the key techniques and challenges. We also organize the milestone works in temporal order so that readers of this survey can comprehend the research roadmap in this field.  Moreover, we also categorized and discussed extended IM problems that extend the classical IM problem with specific real-world features. Lastly, we identified a set intriguing open questions as future directions of research in this arena. Successful solutions to these questions may potentially lead to the fruition of real-world impact of IM research.

\end{document}